\documentclass{article}
\usepackage[utf8]{inputenc}
\usepackage{amsmath}
\usepackage{amssymb}
\usepackage{amsfonts}
\usepackage{amsxtra}
\usepackage{mathrsfs}
\usepackage{graphicx}
\usepackage{subfigure}
\usepackage{caption}
\usepackage{color}
\begin{document}
\newtheorem{algo}{Algorithm}
\newcommand{\krishna}[1]{\textcolor{red}{\textbf{(KG)} #1}}
\newcommand{\greg}[1]{\textcolor{blue}{\textbf{(GT)} #1}}

\title{A graph theoretic framework for representation, exploration and analysis on computed states of physical systems}
\author{R. Banerjee, K. Sagiyama, G.H. Teichert\\Department of Mechanical Engineering\\ \\K. Garikipati\thanks{Corresponding author, {\tt krishna@umich.edu}}\\Departments of Mechanical Engineering \& Mathematics\\ Michigan Insititute for Computational Discovery \& Engineering\\ \\University of Michigan}

\maketitle

\begin{abstract}
    A graph theoretic perspective is taken for a range of phenomena in continuum physics in order to develop representations for analysis of large scale, high-fidelity solutions to these problems. Of interest are phenomena described by partial differential equations, with solutions being obtained by computation. The motivation is to gain insight that may otherwise be difficult to attain because of the high dimensionality of computed solutions. We consider graph theoretic representations that are made possible by low-dimensional states defined on the systems. These states are typically functionals of the high-dimensional solutions, and therefore retain important aspects of the high-fidelity information present in the original, computed solutions.  Our approach is rooted in regarding each state as a vertex on a graph and identifying edges via processes that are induced either by numerical solution strategies, or by the physics. Correspondences are drawn between the sampling of stationary states, or the time evolution of dynamic phenomena, and the analytic machinery of graph theory. A collection of computations is examined in this framework and new insights to them are presented through analysis of the corresponding graphs.
\end{abstract}
\section{Introduction}

In this communication, we explore the casting of large-scale computations of continuum physics in the framework of graph theory. The motivation is to work with low-dimensional representations that encode the fidelity of very high-dimensional, computed solutions, and to develop effective methods to explore and extract further information that could have relevance to decision-making on natural or engineered systems. In this first presentation of ideas, the treatment is deterministic. No probabilistic considerations are invoked.

Of specific interest here are initial and boundary value problems  (IBVPs) that span the range from stationary or steady-state systems through first- and second-order dynamics. Each computation of the IBVPs is at high spatial and/or temporal resolution, making for a high-dimensional and/or long time series numerical solution. While dimensionality reduction techniques such as proper orthogonal decomposition \cite{Sirovich1987,Berkooz1993,Rathnam2003} or tensor decomposition \cite{Tucker1966,Hitchcock1927} methods and their variants have been widely applied to high-dimensional problems, and compelling progress continues to be made on them, our approach is different. With an ultimate view to decision-making on these systems, we consider functionals defined on the high-dimensional numerical solutions, and that are induced by physics on the system-wide scale. Examples of these functionals include the lift, drag and thrust in computational fluid dynamics, the average strain, load at yield, failure strain or dissipated energy in computational solid mechanics, and phase volumes, total and interfacial free energies in computational materials physics. We consider as the state of the physical system, a low-dimensional Euclidean vector (typically of dimension $\sim \mathcal{O}(10^1)$) whose components are such functionals. The crux of our approach is to treat the states as vertices on a graph, which can then be completed by identifying edges between vertex pairs. We seek additionally to introduce edges that are induced by either (a) the numerical solution technique, or (b) a transition guided by some physical property. In the first case the existence of an edge is determined by convergence of a solution step between states. More alternatives present themselves for edge definition via the physics. We present examples that help make these notions more precise. The edges enable graph traversal by time as well as by enumeration of states guided by the numerics or physics in the low-dimensional space. 

Graph theory has a well-appreciated relevance to physical systems, due in some measure to the fact that it underlies network analysis \cite{Newman2010}. Applications including traffic flow, electric power grids and neural circuits are well-known. However, its direct use in representing and analyzing physical phenomena that are described by partial differential equations appears relatively under-explored except for the following work: Graph vertices and edges have been used to represent vortices and their interactions, respectively, for the analysis of turbulent flows in computational fluid dynamics \cite{Nair2015,Taira2016,Scarsoglio2016}. Recently, graphs  have been used to represent problem components such as governing equations, constitutive relations and initial/boundary conditions in the numerical framework of IBVPs \cite{WangSun2018a}. Graph vertices and edges also have been used to represent variables and relations between them, respectively, in a game theoretic approach to discovering constitutive response functions for material failure \cite{WangSun2018b}. 

The following sections consider computations of IBVPs for stationary and steady-state systems (Section \ref{sec:stationary}), non-dissipative dynamics (Section \ref{sec:non-diss-dynamics}) and dissipative dynamics (Section \ref{sec:diss-dynamics}), and connect them to specific types of graphs. The standard machinery of graph theoretic definitions and results is invoked for this purpose. Specific examples are then drawn from computed solutions in a rather extensive Section \ref{sec:computations}. Inferences and insights, not directly available from the high-dimensional numerical solutions, are drawn from analysis of the graphs using well-established concepts from graph theory \cite{West2000,Newman2010}, as well as algorithms that are motivated by the numerical solution or physics specific to each system. Closing remarks are made in Section \ref{sec:closingremarks}.


\section{Stationary and steady-state systems represented as graphs}
\label{sec:stationary}

Consider a domain $\Omega \subset \mathbb{R}^3$, and the following stationary or steady-state BVP:
\begin{subequations}
\begin{alignat}{3}
\nabla\cdot\boldsymbol{\sigma} + \boldsymbol{f} &= \boldsymbol{0}, \quad &&\text{in} \quad &&\Omega\label{eq:stationary-a}\\
\boldsymbol{u} &= \overline{\boldsymbol{u}}(\boldsymbol{p}), \quad &&\text{on} \quad &&\partial\Omega_u\\
\boldsymbol{\sigma}\boldsymbol{n} &= \boldsymbol{\sigma_n}(\boldsymbol{p}), \quad &&\text{on} \quad &&\partial\Omega_\sigma.\label{eq:stationary-c}
\end{alignat}
\end{subequations}

\noindent Here, $\boldsymbol{\sigma}(\boldsymbol{u},\boldsymbol{\alpha};\boldsymbol{p})$ is a (vector or tensor) flux, which is, in general, a functional of its arguments $\boldsymbol{u}$ (the vector of primal variables) and $\boldsymbol{\alpha}$ (the vector of internal state variables), and is parameterized by $\boldsymbol{p}$. We assume that the continuous problem defined by Equations \eqref{eq:stationary-a}--\eqref{eq:stationary-c} is discretized and solved by a numerical method such as finite elements, finite volumes or finite differences, or by Fourier methods among many other possibilities. We refer to this as the numerical problem, and it could be high dimensional in terms of the degrees of freedom (or unknowns). 

States of the system $\mathscr{S}_i \in \mathbb{R}^k$, $i = 1,\dots N$ are low-dimensional vectors with $k$ being much smaller than the dimensionality of the numerical problem. Each $\mathscr{S}_i$ is, in general, a functional of the numerical solution. For stationary or steady state phenomena, different states, $i = 1,\dots N$ are obtained for parameter sets $\boldsymbol{p}_i$, boundary conditions $\overline{\boldsymbol{u}}_i, \boldsymbol{\sigma}_{n_i}$, and boundary decompositions $\partial\Omega = \partial\Omega_{u_i}\cup\partial\Omega_{\sigma_i}$. Also, denote by $\mathscr{T}_{ij}$ a transition from $\mathscr{S}_j$ to $\mathscr{S}_i$ that represents either (a) the nonlinear solution scheme, which, with initial guess $\mathscr{S}_j$, arrives at $\mathscr{S}_i$,\footnote{For elasticity posed as a stationary problem, $\mathscr{S}_j$ and $\mathscr{S}_i$ also admit the added physical interpretations of reference and current, or deformed, states (configurations), respectively. However, for other problems, such as steady heat conduction only the numerical interpretations of initial guess ($\mathscr{S}_j$) and converged solution ($\mathscr{S}_i$) are available.} or (b) a change in a physical quantity between $\mathscr{S}_j$ and $\mathscr{S}_i$,  which we will refer to as a \emph{transition quantity}. To fix ideas, one may consider, as examples of transition quantities, energies and volume fractions of chemical species, among many others. In what follows, we will continue to refer to transition quantities in the abstract, but will make this more specific with examples of graphs in Section \ref{sec:computations}. With this elementary terminology in hand, we can place the description of the states of stationary/steady-state problems and the transitions between them in the graph theoretic setting. For this purpose, we use the standard terminology of graph theory laid out, for instance, by West \cite{West2000}.

\subsection{Vertices, edges and paths; graph properties}
\label{sec:stationary-graphs}

A graph $G(V,E)$ can be constructed such that its set of vertices $V = \{\mathscr{S}_i\}_{i = 1,\dots N}$ and its set of edges $E = \{\mathscr{T}_{ij}\}_{i,j= 1,\dots M}$, for $M \le N$. A number of properties of $G$ can be recognized. Some of them directly reflect definitions of graph theoretic elements. Others are manifestations of the theorems of graph theory and their corollaries. Terminology will be used interchangeably: a vertex for a computed state, an edge for a transition.

\begin{figure}
    \centering
    \includegraphics[scale=0.8]{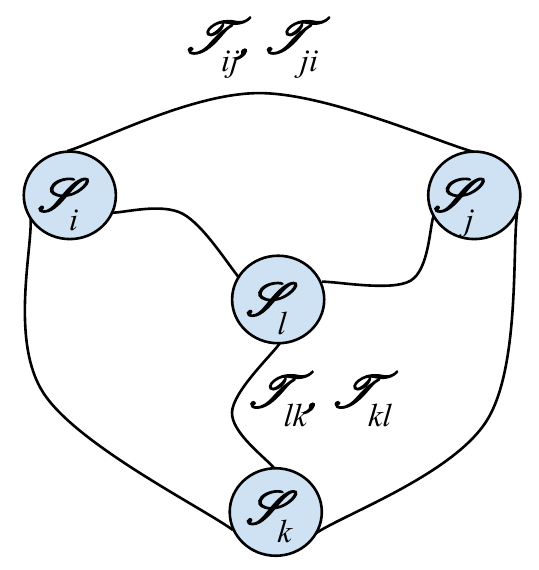}
    \caption{The graph $G(V,E)$, where $V = \{\mathscr{S}_i,\mathscr{S}_j,\mathscr{S}_k,\mathscr{S}_l\}$. Only the edges representing solution schemes/transitions $\{\mathscr{T}_{ij},\mathscr{T}_{ji}\}$ and $\{\mathscr{T}_{lk},\mathscr{T}_{kl}\}$ have been labelled. The graph $G$ is undirected  and is also a clique, representing computations on linear, stationary systems, as well as non-dissipative, dynamical systems under unconditionally stable time integration schemes.}
    \label{fig:statgraph}
\end{figure}

\begin{enumerate}
    \item Graph $G$ is connected if any computed state $\mathscr{S}_i$ must be reached from some initial guess $\mathscr{S}_j$ via a transition $\mathscr{T}_{ij}$ (nonlinear solution step or change in transition quantity). Alternately, certain choices of solution scheme or transition quantity could leave a state $\mathscr{S}_k$ as an unconnected vertex, and the graph $G$ will be unconnected.
    \item The reversibility of linear, non-dissipative systems implies that an edge between $\mathscr{S}_j$ and $\mathscr{S}_i$ can be traversed in either direction: $\mathscr{T}_{ij}$ or $\mathscr{T}_{ji}$. In this sense, linear non-dissipative BVPs are represented by undirected graphs.
    \item The graph $G$ is a clique if it represents a linear, non-dissipative BVP with $E$ defined by numerical solutions between states. In this case, any computed state $\mathscr{S}_i$ can be reached from a starting guess $\mathscr{S}_j$ via the edge $\mathscr{T}_{ij}$. The degree of every vertex is $N-1$: $d_G(\mathscr{S}_i) = N-1,\;\forall i$. The graph $G$ is $(N-1)$-regular.
    \item A vertex $\mathscr{S}_j$ may not be connected to $\mathscr{S}_i$ if the BVP defining the graph is nonlinear and the corresponding states are too distant from each other, with respect to a suitable metric. In such cases, the nonlinear solution scheme may not converge and there may be no edge between $\mathscr{S}_j$ and $\mathscr{S}_i$. The graph $G$ is not a clique in this case. A vertex has degree $d_G(\mathscr{S}_i) \le N-1$. For stiff nonlinear BVPs, the strict inequality is expected to hold. If $E$ is defined by a transition quantity between states, there may not exist a transition to $\mathscr{S}_i$ from $\mathscr{S}_j$ and $G$ is not a clique: $d_G(\mathscr{S}_i) < N-1$.
    \item The above property motivates the definition of an edge weight:
    \begin{equation}
        \text{If}\;\exists \mathscr{T}_{ij},\;\text{then}\quad w_s(\mathscr{T}_{ij}) = \Vert \mathscr{S}_j - \mathscr{S}_i\Vert_V
        \label{eq:weight-s}
    \end{equation}
    where $\Vert\bullet\Vert_V$ is a suitable norm on $V$.
    \item A vertex $\mathscr{S}_i \approx \mathscr{S}$ can be added to $V$ by solving an inverse problem. That is, $\boldsymbol{p}_i$, $\overline{\boldsymbol{u}}(\boldsymbol{p}_i)$, $\boldsymbol{\sigma}_n(\boldsymbol{p}_i)$ can be optimized and $\partial\Omega_{u_i}$ and $\partial\Omega_{\sigma_i}$ chosen so that $\Vert\mathscr{S}_i - \mathscr{S}\Vert < \varepsilon$ for a numerical tolerance $0<\varepsilon \ll 1$.
    \item Typical computations will yield paths on subgraphs of $G$; that is, each state $\mathscr{S}_i$ is only visited once, and the transition (solution step or change in transition quantity)  $\mathscr{T}_{ij}$ is only traversed once. Eulerian tours result if a state $\mathscr{S}_i$ may be revisited, but the transition $\mathscr{T}_{ij}$ is only traversed once, so that subsequent visits to $\mathscr{S}_i$ are from other states $\mathscr{S}_k \neq \mathscr{S}_j$ with transitions $\mathscr{T}_{ik} \neq \mathscr{T}_{ij}$. Cycles follow naturally. 
    \item A branch is created at vertex $\mathscr{S}_i$ if this state can only be reached along edge (transition) $\mathscr{T}_{ij}$, but has two or more solution edges (transitions) leading out from it: $\mathscr{T}_{i_1 i},\mathscr{T}_{i_2 i},\dots$. The states $\mathscr{S}_{i_1},\mathscr{S}_{i_2},\dots$ differ in parameters $\boldsymbol{p}_{i_1},\boldsymbol{p}_{i_2},\dots$, boundary conditions $\{\overline{\boldsymbol{u}}_{i_1}, \boldsymbol{\sigma}_{n_{i_1}}\}, \{\overline{\boldsymbol{u}}_{i_2}, \boldsymbol{\sigma}_{n_{i_2}}\},\dots$, or boundary decompositions $\partial\Omega = \partial\Omega_{u_{i_1}}\cup\partial\Omega_{\sigma_{i_1}} = \partial\Omega_{u_{i_2}}\cup\partial\Omega_{\sigma_{i_2}},\dots$.
\end{enumerate}

Because computed solutions of linear BVPs can attain any state from any other, the corresponding clique graphs are somewhat less interesting. Figure \ref{fig:statgraph} illustrates a clique that could be generated by such  a  system. The graphs become more interesting if some edges are not present, such as due to the choice of transition quantity. The absence of edges (transitions) in systems representing nonlinear BVPs leads to graphs that are not cliques (Property 4). In the absence of a transition between states $\mathscr{S}_j$ and $\mathscr{S}_i$, however, there will typically be a multi-edge path $\mathscr{P}_{ij} = \mathscr{T}_{ik}\cup\mathscr{T}_{kl}\dots\cup\mathscr{T}_{mn}\cup\mathscr{T}_{nj}$. The graph in this case holds information about states that cannot be attained directly from each other, but can be reached via intermediate states. Such examples are presented and analyzed in Sections \ref{sec:nonconvexelasticity}-\ref{sec:DNS-MLgraphs}

\section{Non-dissipative dynamical systems}
\label{sec:non-diss-dynamics}

We consider second-order dynamics in the form of the following IBVP:

\begin{subequations}
\begin{alignat}{3}
\rho \frac{\partial^2\boldsymbol{u}}{\partial t^2} + \nabla\cdot\boldsymbol{\sigma} + \boldsymbol{f} &= \boldsymbol{0}, \quad &&\text{in} \quad &&\Omega\label{eq:sec-order-dyn-a}\\
\boldsymbol{u} &= \overline{\boldsymbol{u}}(\boldsymbol{p}), \quad &&\text{on} \quad &&\partial\Omega_u\label{eq:sec-order-dyn-b}\\
\boldsymbol{\sigma}\boldsymbol{n} &= \boldsymbol{\sigma_n}(\boldsymbol{p}), \quad &&\text{on} \quad &&\partial\Omega_\sigma.\label{eq:sec-order-dyn-c}\\
\boldsymbol{u} &=\boldsymbol{u}_0, \quad &&\text{at} \quad && t = 0\label{eq:sec-order-dyn-d}\\
\dot{\boldsymbol{u}} &=\boldsymbol{v}_0, \quad &&\text{at} \quad && t = 0\label{eq:sec-order-dyn-e}.
\end{alignat}
\end{subequations}
For (non)linear elastodynamics (the prominent example of this type of IBVP), $\boldsymbol{u} \in \mathbb{R}^{n_\text{dim}}$ is the displacement field, its second-order time derivative is the acceleration, with $\rho$ being the mass density. In this case, initial conditions on the primal variable, which is the displacement, and its velocity appear in Equations \eqref{eq:sec-order-dyn-d} and \eqref{eq:sec-order-dyn-e}. For this class of IBVPs, the natural definition of states is for them to be parameterized by time, $t$. The state $\mathscr{S}_i$ is \emph{first} attained at $t = t_i$. 

\subsection{Properties of graphs representing non-dissipative dynamical systems}
\label{sec:non-diss-dynamics-properties}

With IBVPs now in consideration, the notion of initial guess state is replaced by an initial state from which the system evolves dynamically to another state. The edge $\mathscr{T}_{ij}$ represents a single time step between initial state $\mathscr{S}_j$ and final state $\mathscr{S}_i$. Over the time step, a nonlinear (in general) solution step occurs, as well as changes in any transition quantities. The two possible distinct approaches to defining edges thus collapse to the time step. These ideas will be revisited with graph examples in Section \ref{sec:computations}. The following observations can be made regarding the properties introduced for stationary systems:
\begin{enumerate}
    \item Property 1 holds with the replacement of initial guesses with initial states.
    \item The time reversal of non-dissipative dynamical systems, such as second-order elastodynamics, implies that edges can be traversed in either direction between states $\mathscr{S}_j$ and $\mathscr{S}_i$. The graph $G$ is undirected; Property 2 holds.
    \item Assuming unconditionally stable, high-order time integration schemes, all states $\mathscr{S}_i$ admitted by a linear IBVP can be reached by the solution scheme, starting from any other state, $\mathscr{S}_j$. The high-order accuracy is needed to ensure that state $\mathscr{S}_i$ is attained up to the desired tolerance. Property 3 holds; linear IBVPs are represented by cliques. A graph with $N$ vertices is $(N-1)$-regular.
    \item The assumption of an unconditionally stable, high-order integration scheme implies that, even for nonlinear IBVPs (nonlinear elastodynamics, for example), a state $\mathscr{S}_i$ can always be reached in a single time step from initial condition $\mathscr{S}_j$. Property 4 is modified to state that the graph $G$ is a clique even for nonlinear IBVPs.
    \item The definition of an edge weight in Property 5 continues to hold. However, also refer to Property 9 below.
    \item Property 6 on addition of a vertex holds. The numerical stability conferred on the Jacobian of the forward problem by an unconditionally stable scheme also is reflected in the solution of the inverse problem, if solved by adjoint methods.
    \item Property 7 holds; of special interest here are cycles that correspond to periodic solutions, or orbits, of the IBVP.
    \item Property 8 holds unchanged.
\end{enumerate}

The natural parameterization introduced by time leads to additional features in the graphs of non-dissipative dynamical systems.

\begin{itemize}
    \item[9.] Another definition of an edge weight can be introduced as a function of the time between first visits to the given states. Thus,
    \begin{equation}
         w_t(\mathscr{T}_{ij}) = f(\vert t_{i} - t_{j}\vert)
        \label{eq:weight-t}
    \end{equation}
    where it bears emphasis that $t_i$ is the first time that $\mathscr{S}_i$ is attained. Clearly, the existence of periodic orbits implies cycles on the graph, and that $\mathscr{S}_i$ also may be visited at times $\{t^1_i, t^2_i,\dots,t^n_i\}$. This definition of weight collapses to the time step $\Delta t_{ij}$ between first visits to the states $\mathscr{S}_j$ and $\mathscr{S}_i$. Other weight definitions are possible as we will see in Section \ref{sec:DNS-MLgraphs}.
\end{itemize}

 Figure \ref{fig:statgraph} continues to represent the important properties of graphs of non-dissipative dynamical systems. Variations on parameter sets $\boldsymbol{p}$, boundary conditions, boundary decompositions $\partial\Omega = \partial\Omega_{u}\cup\partial\Omega_{\sigma}$, and initial conditions lead to distinct trajectories in the dynamics, each of which can be represented by its own graph.
 
\section{Dissipative dynamical systems}
\label{sec:diss-dynamics}

We consider first-order dynamics in the form of the following IBVP:

\begin{subequations}
\begin{alignat}{3}
\rho \frac{\partial u}{\partial t} + \nabla\cdot\boldsymbol{\sigma} + f &= \boldsymbol{0}, \quad &&\text{in} \quad &&\Omega\label{eq:first-order-dyn-a}\\
u &= \overline{u}(\boldsymbol{p}), \quad &&\text{on} \quad &&\partial\Omega_u\label{eq:first-order-dyn-b}\\
\boldsymbol{\sigma}\boldsymbol{n} &= \boldsymbol{\sigma_n}(\boldsymbol{p}), \quad &&\text{on} \quad &&\partial\Omega_\sigma.\label{eq:first-order-dyn-c}\\
u &=u_0, \quad &&\text{at} \quad && t = 0\label{eq:first-order-dyn-d}.
\end{alignat}
\end{subequations}

\begin{figure}[tb]
        \centering
\begin{minipage}[t]{0.75\textwidth}
        \centering
	\includegraphics[scale=1.0]{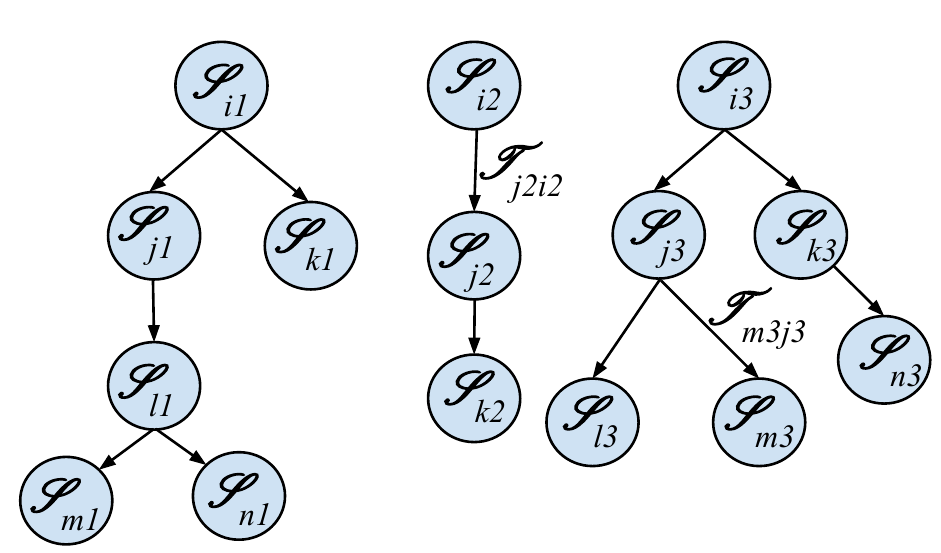}
	\captionof{subfigure}{}
\end{minipage}%
\begin{minipage}[t]{0.25\textwidth}
        \centering
	\includegraphics[scale=1.0]{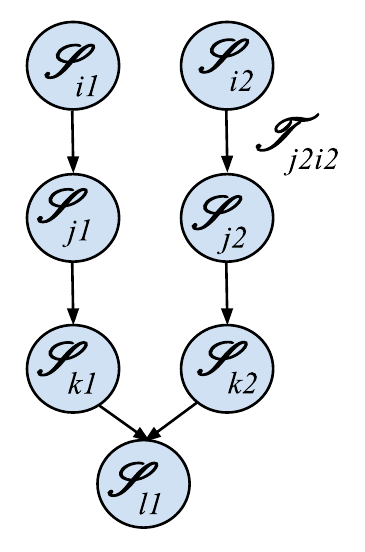}
	\captionof{subfigure}{}
\end{minipage}
        \caption{(a) The graph $G(V,E)$ representing a dissipative dynamical system, where $V = \{\mathscr{S}_{i_1},\mathscr{S}_{j_1},\dots,\mathscr{S}_{i_2},\mathscr{S}_{j_2},\dots,\mathscr{S}_{i_3},\mathscr{S}_{j_3},\dots\}$. The dissipative nature of the dynamical systems represented in this case implies that the existence of edge $\mathscr{T}_{ij}$ means the non-existence of $\mathscr{T}_{ji}$, rendering $G$ a directed graph. If the initial states of sub-graphs $G_1 = (V_1,E_1), G_2 = (V_2,E_2), \dots$ are chosen arbitrarily, then $G$ is, in general an unconnected graph of disjoint trees. See Property 15. (b) It is possible for the graph $G(V,E)$ to contain a vertex that is reached from multiple vertices when representing, for example, an equilibirium state that is attained from multiple initial states.}
	\label{fig:dissipgraph}
\end{figure}

\noindent Note that now, $u \in \mathbb{R}$. For first-order dynamics such as heat conduction and mass transport (of the Fickian, chemical potential-driven, conservative or non-conservative phase field type), $\rho$ is either the heat capacity or equals one, respectively. The flux is of the form $\boldsymbol{\sigma} = -M\nabla u$ for Fourier heat conduction and Fickian diffusion. For conservative phase field models, such as the Cahn-Hilliard equations \cite{CahnHilliard1958}, $\boldsymbol{\sigma} = -M(\nabla\mu(u) -\kappa \nabla\nabla^2 u)$, where $\mu(u)$ is the chemical potential obtained as the derivative of a non-convex function with multiple (at least two) minima, and $\kappa$ represents an interfacial energy. Non-conservative phase field models (Allen-Cahn, \cite{AllenCahn1979}) have $\boldsymbol{\sigma} = -\kappa M\nabla u$, and $f(u)$ is obtained as $M$ multiplied by the derivative of a non-convex function with multiple (at least two) minima. Such examples are presented and explored in Sections \ref{sec:CHdynamicsgraphs} and \ref{sec:DNS-MLgraphs}. Again, different states, $\mathscr{S}_i$, are defined by the time at which they were attained, $t = t_i$. An important aspect of the graphs in this case is that states cannot be revisited, as we point out below. Variations in parameter sets $\boldsymbol{p}_i$, boundary conditions, boundary decompositions $\partial\Omega = \partial\Omega_{u_i}\cup\partial\Omega_{\sigma_i}$, and initial conditions correspond to distinct trajectories, each represented by its own graph, or sub-graph as illustrated in Figure \ref{fig:dissipgraph}.

Dissipative dynamical systems also can be of second order, if the PDE in Equation (\ref{eq:sec-order-dyn-a}) is extended to include a first order term:
\begin{equation}
    \rho \frac{\partial^2\boldsymbol{u}}{\partial t^2} + \gamma \frac{\partial \boldsymbol{u}}{\partial t} + \nabla\cdot\boldsymbol{\sigma} + \boldsymbol{f} = \boldsymbol{0}, \quad \text{in} \quad \Omega\label{eq:sec-order-dissip-dyn}
\end{equation}
Dissipation can be introduced to stationary and steady-state systems such as Equation (\ref{eq:stationary-a}) via an internal variable that is itself governed by first-order dynamics:
\begin{equation}
    \frac{\partial \boldsymbol{\alpha}}{\partial t} = \boldsymbol{g}(\boldsymbol{u},\boldsymbol{\alpha}), \quad \text{in} \quad \Omega\label{eq:dissip-stationary}
\end{equation}
Rate formulations of continuum plasticity fit the description in Equation \eqref{eq:dissip-stationary}, where $\boldsymbol{\alpha}$ would be the equivalent plastic strain.
\subsection{Properties of graphs representing dissipative dynamical systems}
\label{sec:diss-dynamics-properties}

We maintain the convention that $\mathscr{T}_{ij}$ represents a single time step of the computation between states $\mathscr{S}_j$ and $\mathscr{S}_i$. The following observations can be made regarding the properties introduced previously:
\begin{enumerate}
    \item Property 1 holds with a vertex $\mathscr{S}_j$ being the initial state from which $\mathscr{S}_i$ is attained along edge $\mathscr{T}_{ij}$.
    \item The dissipative nature of the systems now being considered means a loss of time reversal symmetry. Consequently, the edge between $\mathscr{S}_j$ and $\mathscr{S}_i$ cannot be traversed in both directions. The graph is directed: if $\exists \mathscr{T}_{ij}, \Longrightarrow \nexists \mathscr{T}_{ji}$. This is a central aspect of graphs representing dissipative dynamical systems, and is illustrated by arrows on directed edges in Figure \ref{fig:dissipgraph}.
    \item Even assuming unconditionally stable, high-order time integration schemes, a state $\mathscr{S}_j$ admitted by a linear IBVP is connected to only certain other states $\{\mathscr{S}_i,\mathscr{S}_k,\dots\}$. The loss of time time reversal symmetry and the broader property of dissipation prevent $G$ from being a clique. This theme is further echoed in Properties 7 and 10-15.
    \item Property 4 on the existence of an edge between a pair of states draws from the dissipative nature of the dynamical system, which confers added stability to integration schemes. States $\mathscr{S}_j$ and $\mathscr{S}_i$ may be distant from each other in a suitable norm, but the existence/non-existence of the edge $\mathscr{T}_{ij}$ also depends on Properties 10-15, induced by the dissipative dynamics, rather than solely on $w_s(\mathscr{T}_{ij})$.
    \item Properties 5 and 9 regarding edge weights of non-dissipative dynamical systems hold for dissipative dynamical systems.
    \item Property 6 on addition of a vertex holds. The added stability from dissipation will be reflected in the solution of inverse problems by adjoint methods to determine states  added as vertices to the graph.
    \item Property 7 suffers a major restriction: Dissipation means that no walk exists that visits the same state $\mathscr{S}_i$ more than once. It follows that the solution step $\mathscr{T}_{ij}$ can only be traversed once. Cycles do not occur. Paths are the only admitted walks.
    \item Property 8 holds unchanged.
    \item Property 9 on time-defined weights holds.
\end{enumerate}

\noindent The following properties are particular to graphs representing dissipative dynamical systems:
\begin{itemize}
    \item[10.] There does not exist a path starting at vertex $\mathscr{S}_j$ and ending at $\mathscr{S}_i$ if the times satisfy $t_i < t_j$.
    \item[11.] If vertex $\mathscr{S}_i$ cannot be reached within a single edge by starting from $\mathscr{S}_j$, then either $t_i < t_j$, or $\Vert\mathscr{S}_j - \mathscr{S}_i\Vert_V > \varepsilon$ for some $\varepsilon > 0$, implying that $\mathscr{S}_i$ is ``distant'' from $\mathscr{S}_j$.
    \item[12.] For the directed graph, the quantities in $\mathscr{S}_j \in \mathbb{R}^k$ can be defined to include at least one whose rates along any directed edge, $\mathscr{T}_{ij}$ are either non-negative or non-positive. These could be observed quantities drawn from $\boldsymbol{u}$ or internal variables drawn from $\boldsymbol{\alpha}$, already governed by Equation (\ref{eq:dissip-stationary}).
    \begin{equation}
        \dot{f}(\boldsymbol{u}) \ge 0,\quad\text{or}\; \dot{f}(\boldsymbol{u}) \le 0, \qquad \dot{\alpha}^A \ge 0,\quad\text{or}\; \dot{\alpha}^A \le 0,
        \label{eq:entropy-quant}
    \end{equation}
    where $A$ denotes the components of $\boldsymbol{\alpha}$. We will use the term \emph{entropy quantities} for $f(\boldsymbol{u})$ and $\alpha^A$ of the type in Equation \eqref{eq:entropy-quant} by analogy with the non-negativity of the entropy rate. 
    \item[13.] Note that the states that define the vertices of $G$ must contain entropy quantities if the graph is to properly represent a dissipative dynamical process.
    \item[14.] Dissipation implies that all walks are paths, and therefore no cycles are allowed. It follows, then, that these graphs are trees \cite{West2000}.
    \item[15.] A tree, $G_1(V_1,E_1)$, has as its root an initial state $\mathscr{S}_{i_1}$, and remains disjoint from the tree, $G_2(V_2,E_2)$, with root $\mathscr{S}_{i_2}$, unless the vertices on $G_1$ and $G_2$ are chosen to ensure that the entropy quantities are identical for some states $\mathscr{S}_{j_1} \in V_1$ and $\mathscr{S}_{j_2} \in V_2$. The graph $G = G_1 \cup G_2 \cup \dots$ is not, in general, fully-connected, as illustrated in Figure \ref{fig:dissipgraph}a. However, Figure \ref{fig:dissipgraph}b illustrates a situation in which the sub-trees $G_1$ and $G_2$ have been connected, by choosing vertices $\mathscr{S}_{k_1}$ and $\mathscr{S}_{k_2}$, and the transitions $\mathscr{T}_{{l_1}{k_1}}$ and $\mathscr{T}_{{l_1}{k_2}}$ so that a branch exists at vertex $\mathscr{S}_{l_1}$.
\end{itemize}

\section{Graph theoretic representations of computed solutions}
\label{sec:computations}
We consider examples that span the systems outlined abstractly in Sections \ref{sec:stationary}--\ref{sec:diss-dynamics}. In each case, we present an outline of the partial differential equations and computational physics frameworks followed by discussions of the graphs induced by the computed numerical solutions.

\subsection{Graphs constructed on solutions to non-dissipative elastodynamics and linear elasticity}
\label{sec:linelasticity}

\begin{figure}
    \centering
    \includegraphics[scale=1.0]{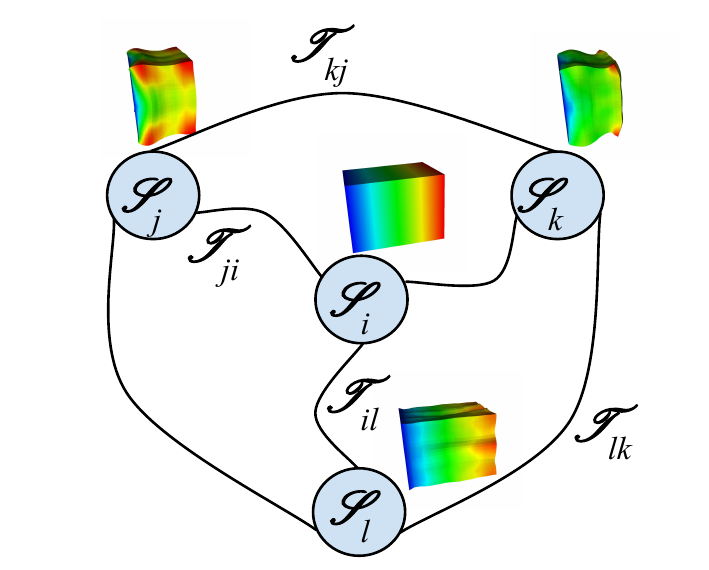}
    \caption{A clique representing either a non-dissipative dynamical system of elastodynamics computed with an unconditionally stable, second-order accurate solution scheme, or a stationary problem of linear, quasistatic elasticity. The deformation has been scaled $40\times$ for clarity.}
    \label{fig:stat-non-dissip-comput-graph}
\end{figure}

The graph in Figure \ref{fig:stat-non-dissip-comput-graph} is a clique representing a non-dissipative elastodynamics problem. Each vertex shows a computed state solved with an unconditionally stable, second-order scheme, which ensures that each state is attainable from any other. The superposed field represents the Euclidean norm of the displacement, $\boldsymbol{u}$. The state $\mathscr{S}_i$ represents the initial condition, and the cycle $\mathscr{S}_i \rightarrow\mathscr{S}_j \rightarrow\mathscr{S}_k\rightarrow\mathscr{S}_l\rightarrow\mathscr{S}_i$ is a complete period; represented by a cycle on the graph. The edges, $\mathscr{T}_{ji}, \dots \mathscr{T}_{il}$, all have the same time-based weights $w_t$ defined in Equation \eqref{eq:weight-t}. However, their solution norm-based weights $w_s$, defined via Equation \eqref{eq:weight-s}, will differ. This graph is isomorphic to one in which the vertices represent stationary states of deformation of a quasi-statically strained, linearly elastic solid. The computed states, of course, would be different from those in the figure. In such a stationary problem of linearized elasticity, any state can be attained from any other state, thus maintaining the cliqueness of the graph. The edges now represent solution steps between states, and admit only the solution norm-based weights, $w_s$. The graph is undirected, reflecting the time-reversal symmetry of elastodynamics, Equation \eqref{eq:sec-order-dyn-a}, and the reversibility of the stationary problem of elasticity, Equation \eqref{eq:stationary-a}. Both these properties are inherited by the respective numerical schemes: an unconditionally stable, second-order accurate method for the elastodynamics problem, and a linear solve for linearized elasticity. In both these systems of equations the constitutive relation $\boldsymbol{\sigma} = \mathrm{sym}[\partial \hat{\psi}/\partial\nabla^\text{s}\boldsymbol{u}]$ holds for a strain energy density function $\hat{\psi}(\nabla^\text{s}\boldsymbol{u})$, where $\nabla^\text{s}$ is the symmetric gradient.

\subsection{Graphs on stationary states of gradient-regularized, non-convex elasticity at finite strain}

\label{sec:nonconvexelasticity}

\begin{figure}
    \centering
    \includegraphics[scale=0.3]{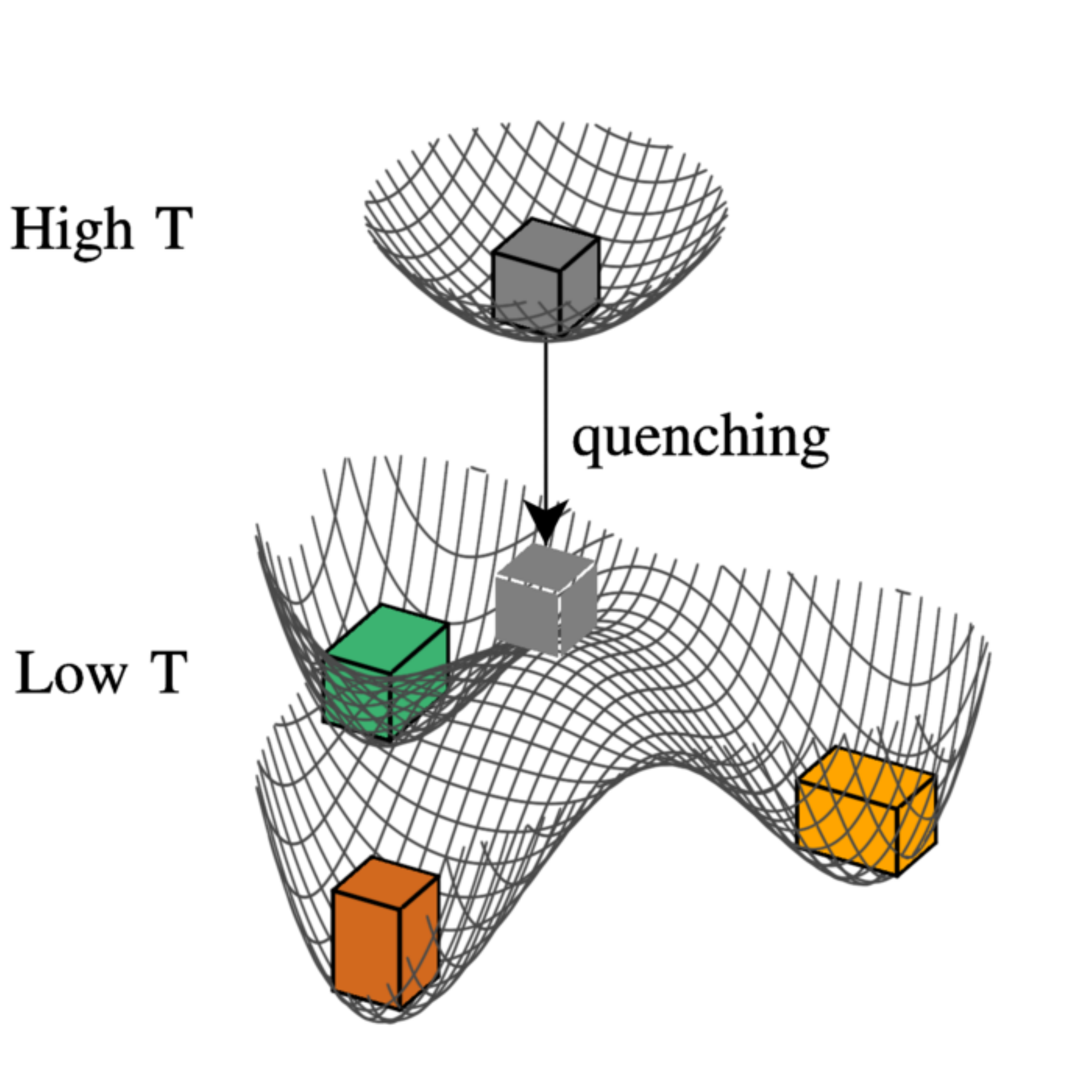}
    \caption{Illustration of the free energy density component, $\widetilde{\psi}$, as a surface undergoing a convex to non-convex transition in a suitably parameterized, reduced strain space. The parameterization does not appear explicitly, but is represented by the transformation from a cubic crystal structure to three tetragonal structures with the same symmetry group.}
    \label{fig:freeenergy1}
\end{figure}

We next consider graphs induced by the states that arise as free energy minima \cite{Rudrarajuetal2016,SagiyamaGarikipati2017a,SagiyamaGarikipati2017b} in a gradient-regularized model of non-convex elasticity at finite strain \cite{Toupin1962,Barsch1984}. In summarizing the problem we  begin with the free energy density function, now written as $\bar{\psi}(F_{iJ},F_{iJ,K})$, where, with $u_i$ denoting the displacement vector, and $X_I$ denoting the reference position in coordinate notation, the deformation gradient tensor $F_{iJ}$ and its gradient $F_{iJ,K}$ are

\begin{equation}
    F_{iJ} = \delta_{iJ} + \frac{\partial u_i}{\partial X_J}; \qquad F_{iJ,K} = \frac{\partial F_{iJ}}{\partial X_K}
    \label{eq:defgrad}
\end{equation}
As illustrated in Figure \ref{fig:freeenergy1} the solid undergoes a transition from a high temperature cubic crystal structure, in which the free energy density $\bar{\psi}$ is fully convex, to a low temperature tetragonal crystal structure, in which a non-convex component appears in $\bar{\psi}$. We focus on describing the response of the solid in this regime, where the non-convex component, $\widetilde{\psi}$ must be accounted for in an appropriate constitutive description  by writing it as a function of suitably parameterized strain quantities. For details of the parameterization the interested reader is directed to the the work of Barsch \& Krumhansl \cite{Barsch1984}. 

The non-convexity in $ \widetilde{\psi}$ gives rise to finely fluctuating solution fields. They model martensitic microstructures, which can be solved for if the free energy density is coercified by a strain gradient contribution:
\begin{equation}
    \bar{\psi}(F_{iJ}, F_{iJ,K}) = \widetilde{\psi}(F_{iJ}) + \lambda F_{iJ,K}F_{iJ,K}.
    \label{eq:freeenergy}
\end{equation}
 where $\lambda$ is related to the energy of interfaces. 

The stress and higher-order stress, respectively, are then written as:

\begin{subequations}
\begin{align}
 P_{iJ} &= \frac{\partial \bar{\psi}}{\partial F_{iJ}}\label{eq:stressP} \\
 B_{iJK} &= \frac{\partial \bar{\psi}}{\partial F_{iJ,K}}\label{eq:stressB}
\end{align}
\end{subequations}

The governing system of partial differential equations is:

\begin{subequations}
\begin{alignat}{2}
P_{iJ,J} - B_{iJK,JK} &= 0 &&\mathrm{in} ~\Omega\label{eq:strongformgradelasticity-a}\\
u_{i}  &= \bar{u}_i   &&\mathrm{on} ~\partial\Omega_{i}^u\label{eq:strongformgradelasticity-b}\\
P_{iJ}N_J - DB_{iJK}N_KN_J - 2D_J(B_{iJK})N_K &  &&\phantom{.} \nonumber\\
- B_{iJK}D_JN_K + (b^L_LN_JN_K-b_{JK})B_{iJK} & = T_{i} &&\mathrm{on} ~\partial\Omega_{i}^T\label{eq:strongformgradelasticity-c}\\
Du_i  &= 0 &&\mathrm{on} ~\partial\Omega_{i}^m\label{eq:strongformgradelasticity-d}\\
B_{iJK}N_JN_K &= 0 &&\mathrm{on}  ~\partial\Omega_{i}^M\label{eq:strongformgradelasticity-e}
\end{alignat}
\end{subequations}

\noindent  where, $\partial\Omega= \partial\Omega_{i}^u \cup  \partial\Omega_{i}^T$ and $\partial\Omega= \partial\Omega_{i}^m \cup  \partial\Omega_{i}^M$ represent distinct decompositions of the smooth boundary. Here, $b_{IJ}=-D_{I}N_J=-D_{J}N_I$ are components of the second fundamental form of the smooth boundary. The governing partial differential equation in \eqref{eq:strongformgradelasticity-a} is in conservation form of \eqref{eq:stationary-a}, which can be seen by identifying $\sigma_{iJ} = P_{iJ} - B_{iJK,K}$. It also is nonlinear and fourth-order as is apparent upon substituting Equations \eqref{eq:defgrad}, \eqref{eq:freeenergy} and \eqref{eq:stressB} for $B_{iJK,JK}$ in Equation \eqref{eq:strongformgradelasticity-a}. The Dirichlet boundary condition in \eqref{eq:strongformgradelasticity-b} has the same form as for conventional elasticity. Equation \eqref{eq:strongformgradelasticity-c} is the extension of \eqref{eq:stationary-c}, while \eqref{eq:strongformgradelasticity-d} and \eqref{eq:strongformgradelasticity-e} are additional requirements, all arising from the fourth-order differential nature of this problem.\footnote{The Neumann boundary condition, \eqref{eq:strongformgradelasticity-c} is notably more complex than its conventional counterpart, which would have only the first term on the left hand-side. Equation \eqref{eq:strongformgradelasticity-d} is the higher-order Dirichlet boundary condition applied to the normal gradient of the displacement field, and Equation \eqref{eq:strongformgradelasticity-e} is the higher-order Neumann boundary condition on the higher-order stress, $\boldsymbol{B}$. Adopting the physical interpretation of $\boldsymbol{B}$ as a couple stress \cite{Toupin1962}, the homogeneous form of this boundary condition, if extended to the atomic scale, states that there is no boundary mechanism to impose a generalized moment across atomic bonds.}

\begin{figure}
    \centering
    \includegraphics[scale=0.3]{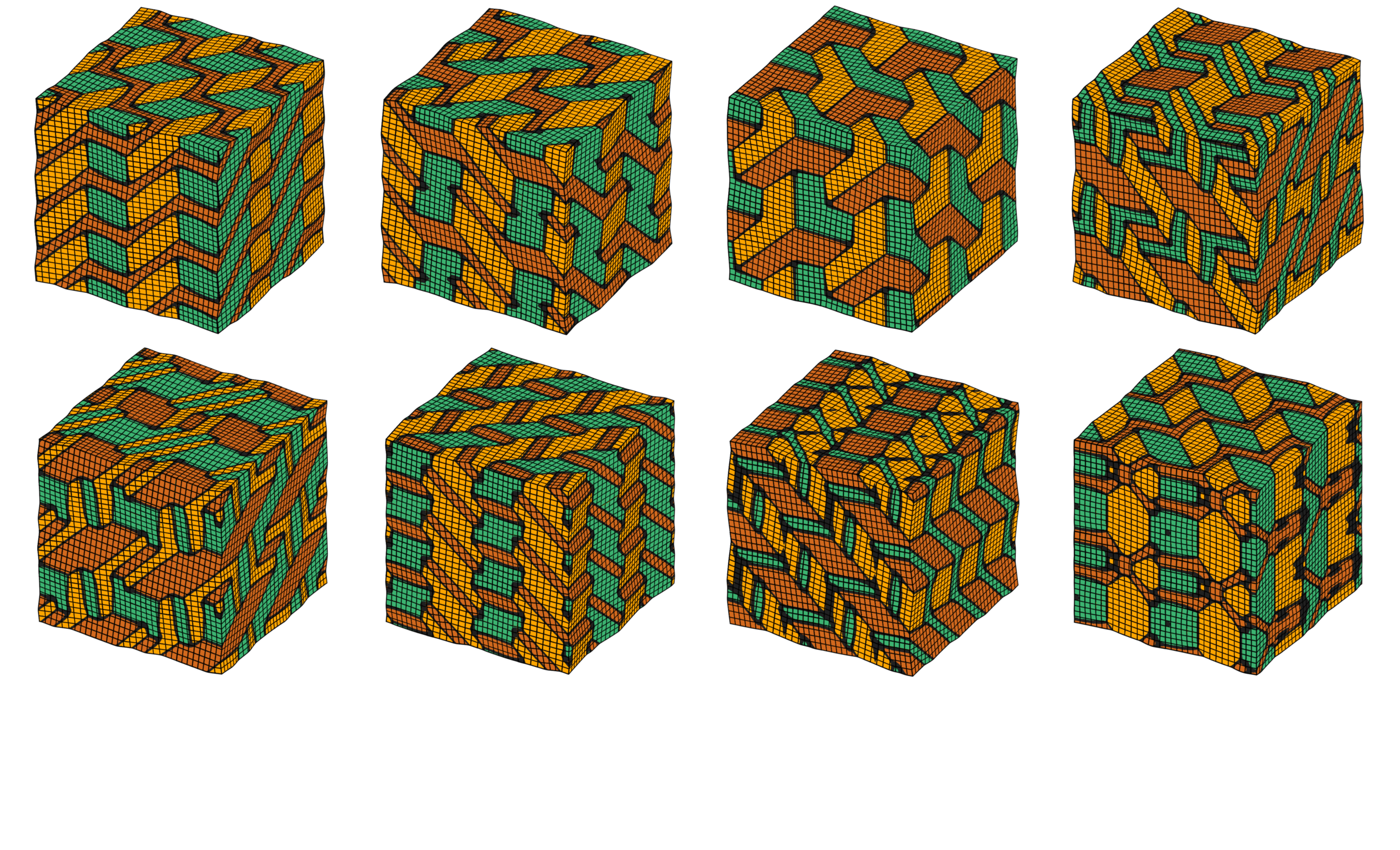}
    \caption{Example microstructures obtained with the nonconvex model of elasticity around a mean deformation gradient $\frac{1}{\text{meas}(\Omega)}\int_\Omega \boldsymbol{F}\mathrm{d}V = \boldsymbol{1}$.}
    \label{fig:microstructures}
\end{figure}

With the mathematical formulation of the problem outlined above, we have computed a number of martensitic microstructures, some of which appear in Figure \ref{fig:microstructures}. The three tetragonal crystal structures in Figure \ref{fig:freeenergy1} occur in the martensitic microstructures of Figure \ref{fig:microstructures}, with each tetragonal variant being represented by the same color in the two figures. A more detailed exposition of this problem from the mathematical and computational points of view has been laid out elsewhere \cite{Rudrarajuetal2014,Rudrarajuetal2016,Sagiyamaetal2016,SagiyamaGarikipati2017b}. Using isogeometric analytic methods, described in the preceding references, for the ease of representing finite-dimensional functions of high-order continuity, we have obtained numerical solutions to boundary value problems posed on the system of equations (\ref{eq:defgrad}-\ref{eq:strongformgradelasticity-e}). With periodic boundary conditions, we have obtained a range of microstructures that appear in Figure \ref{fig:microstructures}.

 Because of the differing orientations of the tetragonal variants, the elastic response fluctuates rapidly over each microstructure shown in Figure \ref{fig:microstructures}. In solids of interest for materials physics applications (e.g., batteries, electronics and structural alloys) the sub-domains with a uniform variant are on the scale of microns or less, and it is often of interest to model their effective properties; for instance, the effective stress-strain response. With the aim of carrying out such homogenization numerically (a study that will be described in detail elsewhere), we have computed $2770$ elastic states for the microstructure appearing in the upper left of Figure \ref{fig:microstructures}. These strains were imposed with Dirichlet boundary conditions \eqref{eq:strongformgradelasticity-b} and \eqref{eq:strongformgradelasticity-d} that define a bijective mapping to a set of uniform deformation gradients\footnote{Of course, because of the microstructure of martensitic variants, and the anisotropy induced by it, the actual deformation gradient, $\boldsymbol{F}$ is non-uniform, but satisfies $\frac{1}{\text{meas}(\Omega)}\int_\Omega \boldsymbol{F}\text{d}V = \widehat{\boldsymbol{F}}$.} denoted by $\widehat{\boldsymbol{F}}$, from which the Green-Lagrange strain is obtained as $\widehat{\boldsymbol{E}} = \frac{1}{2}(\widehat{\boldsymbol{F}}^\mathrm{T}\widehat{\boldsymbol{F}} - \boldsymbol{1})$. Because of its symmetry, each tensor-valued $\widehat{\boldsymbol{E}}$ is a point in $\mathbb{R}^6$. 

\begin{figure}
    \centering
    \includegraphics[scale=0.3]{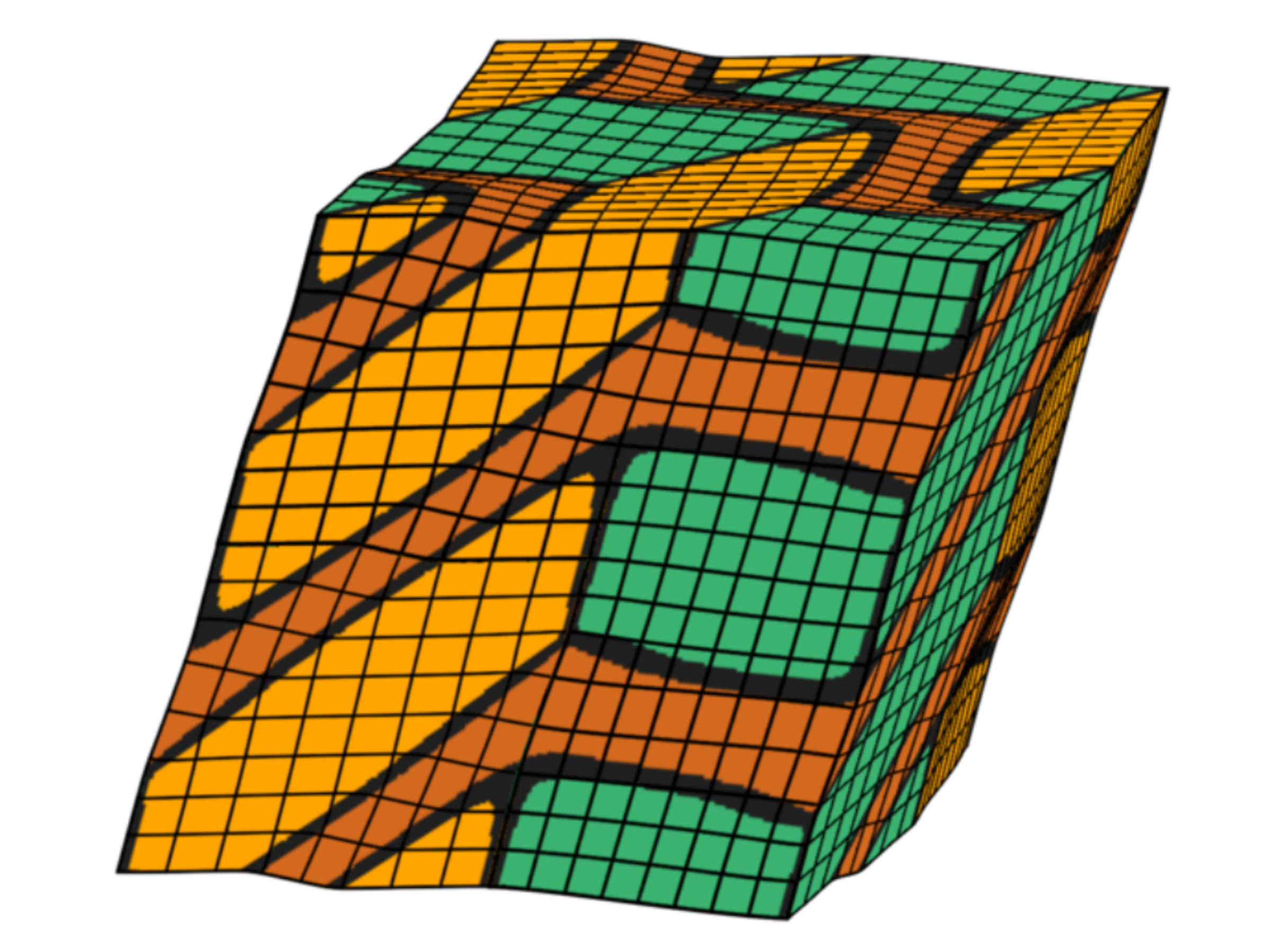}
    \caption{An example of a strain state $\mathscr{S}_i \equiv\widehat{\boldsymbol{E}}_i$ imposed on the microstructure appearing in the upper left of Figure \ref{fig:microstructures}.}
    \label{fig:micro_deform}
\end{figure}

\subsubsection{Graphs induced by strain states}
\label{sec:graphsinducedstrains}
The $N = 2770$ strain states, which we use to parameterize the set of boundary conditions, are labelled by $\widehat{\boldsymbol{E}}_i \in \mathbb{R}^6,\; i = 1,\dots N$. Identifying each state with a graph vertex, we formally write $\mathscr{S}_i \equiv \widehat{\boldsymbol{E}}_i$. These states are chosen from a larger set of states (vertices), $V = \{ \mathscr{S}_i\}_{i=1,\dots M} \equiv \{\widehat{\boldsymbol{E}}_i\}_{i=1,\dots M}$, where $M > N$, of states whose tensor components are generated by a Sobol' sequence, chosen for its space-filling property, in $\mathbb{R}^6$. Starting with an arbitrary strain state, $\mathscr{S}_j \equiv \widehat{\boldsymbol{E}}_j$ from the set $V$, we apply the boundary conditions that define a uniform deformation gradient $\widehat{\boldsymbol{F}}_j$ (unique up to rotations) relative to an undeformed reference state of the crystal occupying region $\Omega \subset \mathbb{R}^3$. The corresponding stationary elastic state of the chosen microstructure is thus computed and satisfies $\frac{1}{\text{meas}(\Omega)}\int_\Omega\boldsymbol{F}\mathrm{d}V = \widehat{\boldsymbol{F}}_j$. We introduce another set $\widetilde{V}$, which will contain those states whose elastic response has been solved for, and initialize it with the state $\mathscr{S}_j \equiv \widehat{\boldsymbol{E}}_j$. So, $\widetilde{V} = \{\mathscr{S}_j \}$. Next, we search for another state, $\mathscr{S}_k \equiv \widehat{\boldsymbol{E}}_k$, subjected to Dirichlet boundary conditions corresponding to the uniform deformation gradient $\widehat{\boldsymbol{F}}_k$ (unique up to rotations). For many choices of $\widehat{\boldsymbol{E}}_k$, the stiffness induced by the microstructure will render the preceding boundary value problem too stiff for convergence of the nonlinear solver. However, if  $\widehat{\boldsymbol{E}}_k$ is within an $\varepsilon$-ball of $\widehat{\boldsymbol{E}}_j$ in the Frobenius norm $\Vert\bullet\Vert_\text{F}$, the relative deformation gradient $\widehat{\boldsymbol{F}}_k\widehat{\boldsymbol{F}}_j^{-1}$ applied to the elastic state $\mathscr{S}_j \equiv \widehat{\boldsymbol{E}}_j$ allows the computation of the elastic state labelled by $\mathscr{S}_k \equiv \widehat{\boldsymbol{E}}_k$. This step induces the transition in states $\mathscr{S}_j\rightarrow \mathscr{S}_k$, i.e., the edge $\mathscr{T}_{kj}$. We expand $\widetilde{V}$ to $\widetilde{V} = \{\mathscr{S}_j,\mathscr{S}_k \}$, and initialize $\widetilde{E} = \{\mathscr{T}_{kj}\}$. In subsequent steps, we search for solutions to the elastic states corresponding to strains drawn from $V$, with initial (reference) states drawn from $\widetilde{V}$. If the initial state is $\mathscr{S}_l \equiv \widehat{\boldsymbol{E}}_l$, and the target state is $\mathscr{S}_m \equiv \widehat{\boldsymbol{E}}_m$, the relative deformation gradient is $\widehat{\boldsymbol{F}}_m\widehat{\boldsymbol{F}}_l^{-1}$ and the transition between states is $\mathscr{S}_l\rightarrow \mathscr{S}_m$, allowing an expansion to $\widetilde{V} = \{\mathscr{S}_j,\mathscr{S}_k,\dots,\mathscr{S}_l,\mathscr{S}_m\}$, and the set of edges to $\widetilde{E} = \{\mathscr{T}_{kj}, \dots,\mathscr{T}_{ml} \}$. The induced graph is $\widetilde{G} = (\widetilde{V},\widetilde{E})$, with $\widetilde{V} \subset V, \widetilde{E} \subset E$. 

If $\varepsilon$ is small enough we find that the nonlinear solver converges for the transitions $\mathscr{S}_l\rightarrow \mathscr{S}_m$ and $\mathscr{S}_m\rightarrow \mathscr{S}_l$, making the edges $\mathscr{T}_{ml}$ undirected. For a given $\varepsilon$, the nonlinear solution may, however, fail to converge for some pairs $\{\mathscr{S}_l,\mathscr{S}_m\}$. If the elastic state $\mathscr{S}_m$ was sought, then it is not added to $\widetilde{V}$, and the edge $\mathscr{T}_{ml}$ is not added to $\widetilde{E}$.\footnote{We found that if the nonlinear solver fails to converge for transition $\mathscr{S}_l\rightarrow \mathscr{S}_m$, it also fails for  $\mathscr{S}_m\rightarrow \mathscr{S}_l$.Then, both the edges $\mathscr{T}_{ml}$ and  $\mathscr{T}_{lm}$ do not exist.} The generation of the induced graph is laid out in Algorithm \ref{algo:strainstates}.

\begin{algo}
Graph generation induced by strain states.
\begin{center}
    \fbox{\begin{minipage}{11cm}
{\tt
\begin{enumerate}
    \item Setup: $V = \{\mathscr{S}_i\}_{i = 1,\dots,M}$, where $\mathscr{S}_i \equiv \widehat{\boldsymbol{E}}_i$; choose $\varepsilon > 0$.
    \item Initialization: $\widetilde{V} = \emptyset$; $\bar{V} = \emptyset$; $\widetilde{E} = \emptyset$
    \item Exploration: while solver\_does\_not\_converge == True 
    \begin{itemize}
        \item[] choose $\mathscr{S}_j \in V\backslash \bar{V}$; 
        \item[] search for the nonlinear solution with boundary conditions corresponding to $\widehat{\boldsymbol{F}}_j$;
        \item[] $\bar{V} = \bar{V}\oplus \{\mathscr{S}_j\}$ (state $\mathscr{S}_j$ could not be solved for);
    \end{itemize}
     
    \item Graph growth: $\widetilde{V} = \widetilde{V} \oplus \{\mathscr{S}_j\}$ (solution to state $\mathscr{S}_j$ converged); 
    
    \item while $\vert \widetilde{V} \vert \le N$;
    \begin{enumerate}
        \item Reinitialization: $\bar{V} = \emptyset$;
        \item Exploration: while solver\_does\_not\_converge == True
        \begin{itemize}
            \item[] choose $\mathscr{S}_l \in \widetilde{V}$;
            \item[] choose $\mathscr{S}_m \in V\backslash \bar{V}$ s.t. $0 < \Vert \widehat{\boldsymbol{E}}_l - \widehat{\boldsymbol{E}}_m \Vert_\text{F} \le \varepsilon$;
            \item[] search for the nonlinear solution with boundary conditions corresponding to $\widehat{\boldsymbol{F}}_m\widehat{\boldsymbol{F}}_l^{-1}$;
            \item[] $\bar{V} = \bar{V}\oplus \{\mathscr{S}_m\}$ (state $\mathscr{S}_m$ could not be solved for);
        \end{itemize}
        \item Graph growth: 
        \begin{itemize}
            \item[] $\widetilde{V} = \widetilde{V} \oplus \{\mathscr{S}_m\}$ (solution to state $\mathscr{S}_m$ converged);
            \item[] $\widetilde{E} = \widetilde{E} \oplus \{\mathscr{T}_{ml}\}$;
        \end{itemize}
         
    \end{enumerate}
\end{enumerate}}
\end{minipage}}
\end{center}
\label{algo:strainstates}
\end{algo}

\begin{figure}
    \centering
    \includegraphics[scale=0.3]{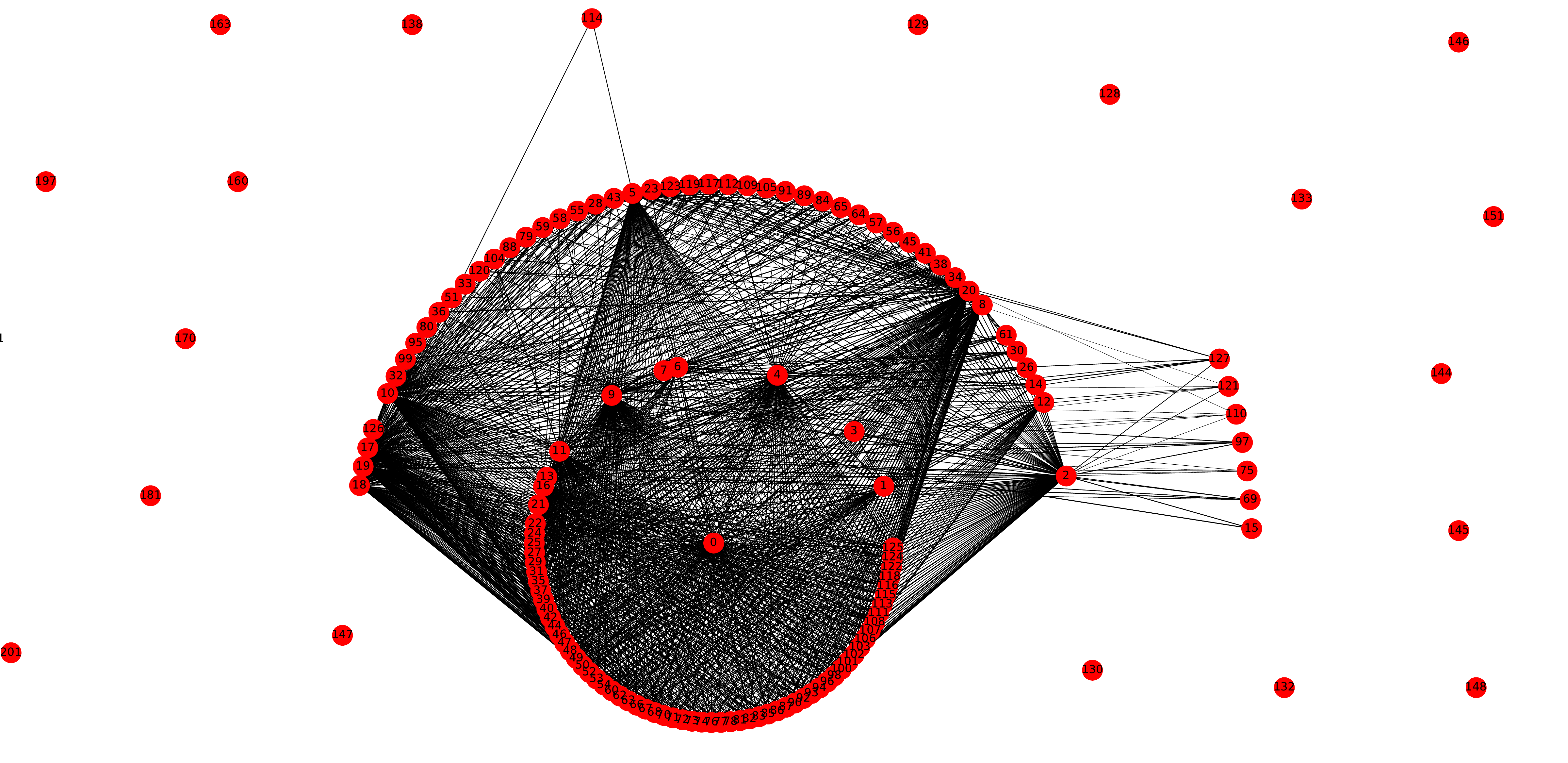}
    \caption{Generation of the graph of strain states by Algorithm \ref{algo:strainstates}, showing a central, densely connected component in circular layout and a few unconnected vertices whose edges remain to be determined at the stage shown. This graph was generated by exploring the strain states of the microstructure in the upper left corner of Figure \ref{fig:microstructures}.}
    \label{fig:microstructureStrains}
\end{figure}

We have applied Algorithm \ref{algo:strainstates} to generate graphs as large as $\vert\widetilde{V}\vert = 2770$. A circular layout of one of the graphs induced by strain states of the microstructure in the upper left corner of Figure \ref{fig:microstructures} during its generation by Algorithm \ref{algo:strainstates} appears in Figure \ref{fig:microstructureStrains}. For clarity and ease of viewing, we have dispensed with the $\mathscr{S},\mathscr{T}$ labelling, and only have indicated vertices (states) by their numbers. Because of the density of edges in larger graphs, we have restricted ourselves, for purposes of illustration, to analyzing a graph of size $\vert\widetilde{V}\vert = 128$ in the remainder of Section \ref{sec:nonconvexelasticity}. 

\subsubsection{Graph layouts; eigenvector centrality and degree centrality reveal the importance of individual strain states to graph traversal}
\label{sec:elastgraphlayouts}

The high degrees of most vertices makes the circular layout in Figure \ref{fig:microstructureStrains} unsuitable for visualization. For this reason we also have presented the graph in the Kamada-Kawai layout \cite{Kamada1989}, which is based on finding local minima of an edge length-dependent energy defined on the two dimensional planar graph. This energy is written as \cite{Koren2005}
\begin{equation}
    U = \sum\limits_{\mathscr{T}_{ij}\in \widetilde{E}}w(\mathscr{T}_{ij})d^2(\mathscr{T}_{ij}) = \sum\limits_{p=1}^2 \boldsymbol{x}_p^\text{T}\boldsymbol{L} \boldsymbol{x}_p
    \label{eq:graphenergy}
\end{equation}
where $d(\mathscr{T}_{ij})$ is the length of edge $\mathscr{T}_{ij}$ in the planar layout of the graph, $\boldsymbol{x}_p \in \mathbb{R}^N$ with $N = \vert\widetilde{V}\vert$, is the vector of the $p^\text{th}$ coordinate of the vertices ($p = 1,2$), and $\boldsymbol{L}$ is the Laplacian matrix of $\widetilde{G}$ \cite{Newman2010}. This matrix is constructed from the adjacency matrix, $\boldsymbol{A}$, and the degree matrix, $\boldsymbol{D}$ as $\boldsymbol{L} = \boldsymbol{D} - \boldsymbol{A}$, where $\boldsymbol{A}$ and $\boldsymbol{D}$ are defined as
\begin{equation}
    A_{ij} = \begin{cases}
                w(\mathscr{T}_{ij}) &\text{if}\; \exists\, \mathscr{T}_{ij}\\
                0 &\text{otherwise}
                \end{cases},\quad 
                D_{ij} = \begin{cases}\sum\limits_{k=1}^N A_{ik}&\text{if}\; i = j\\
                0 &\text{otherwise}
                \end{cases}
\end{equation}
For unweighted matrices, the same definitions work with $w(\mathscr{T}_{ij}) = 1$ if $\exists\;\mathscr{T}_{ij}$. Note that $\boldsymbol{A}$ is symmetric for undirected graphs, such as we have for this problem, and $\boldsymbol{D}$ is diagonal. The minimization problem gives
\begin{equation}
    (\boldsymbol{x}^\text{min}_1,\boldsymbol{x}^\text{min}_2) = \text{arg}\;\min\limits_{(\boldsymbol{x}_1,\boldsymbol{x}_2)} \sum\limits_{p=1}^2 \boldsymbol{x}_p^\text{T}\boldsymbol{L} \boldsymbol{x}_p
    \label{eq:graphenergymin}
\end{equation}

\begin{figure}
    \centering
    \includegraphics[scale=0.4]{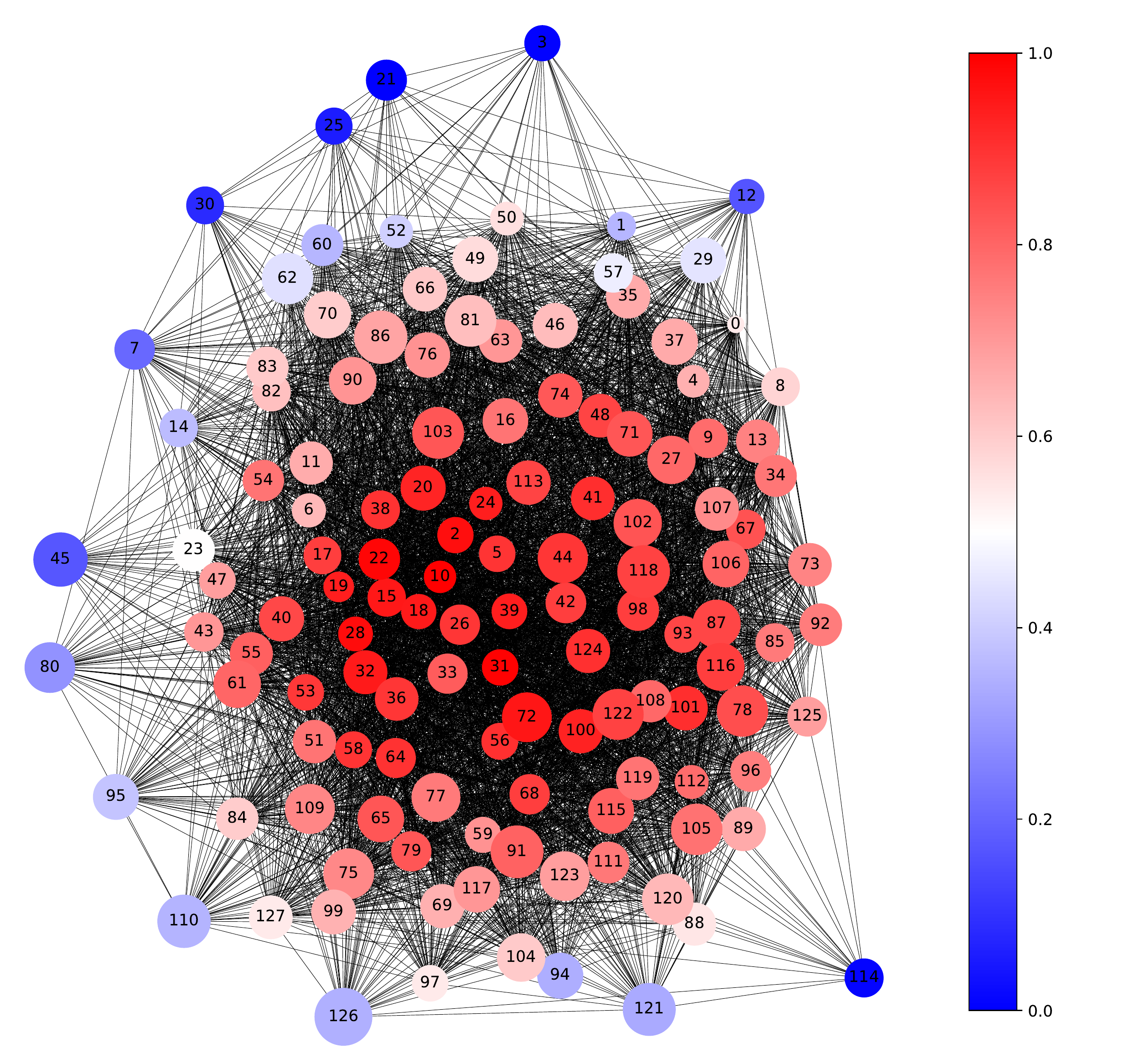}
    \caption{The graph of strain states in Kamada-Kawai layout. The vertices are shaded by their eigenvector centrality, while the area of each vertex symbol is proportional to the Frobenius norm of the strain $\Vert\widehat{\boldsymbol{E}}_i\Vert_\text{F}$.}
    \label{fig:ms_kamadakawai}
\end{figure}

The Kamada-Kawai layout in Figure \ref{fig:ms_kamadakawai} is favored for the visual appeal achieved by energy minimization as defined by Equation \eqref{eq:graphenergymin}. Each vertex symbol has its area scaled by $\Vert\widehat{\boldsymbol{E}}\Vert_\text{F}$. We also computed the eigenvector centrality of $\widetilde{G}$, to elucidate the extent to which a vertex and its associated strain state are connected to (have reversible strain paths to) other states, or to other well-connected states, or both  \cite{Newman2010}. This is a measure of the importance of a strain state (vertex) to navigation of paths through the entire set $\widetilde{V}$ (alternately, the graph) generated by Algorithm \ref{algo:strainstates}. The eigenvector centrality is represented by the shading of vertices. 

\begin{figure}
    \centering
    \includegraphics[scale=0.4]{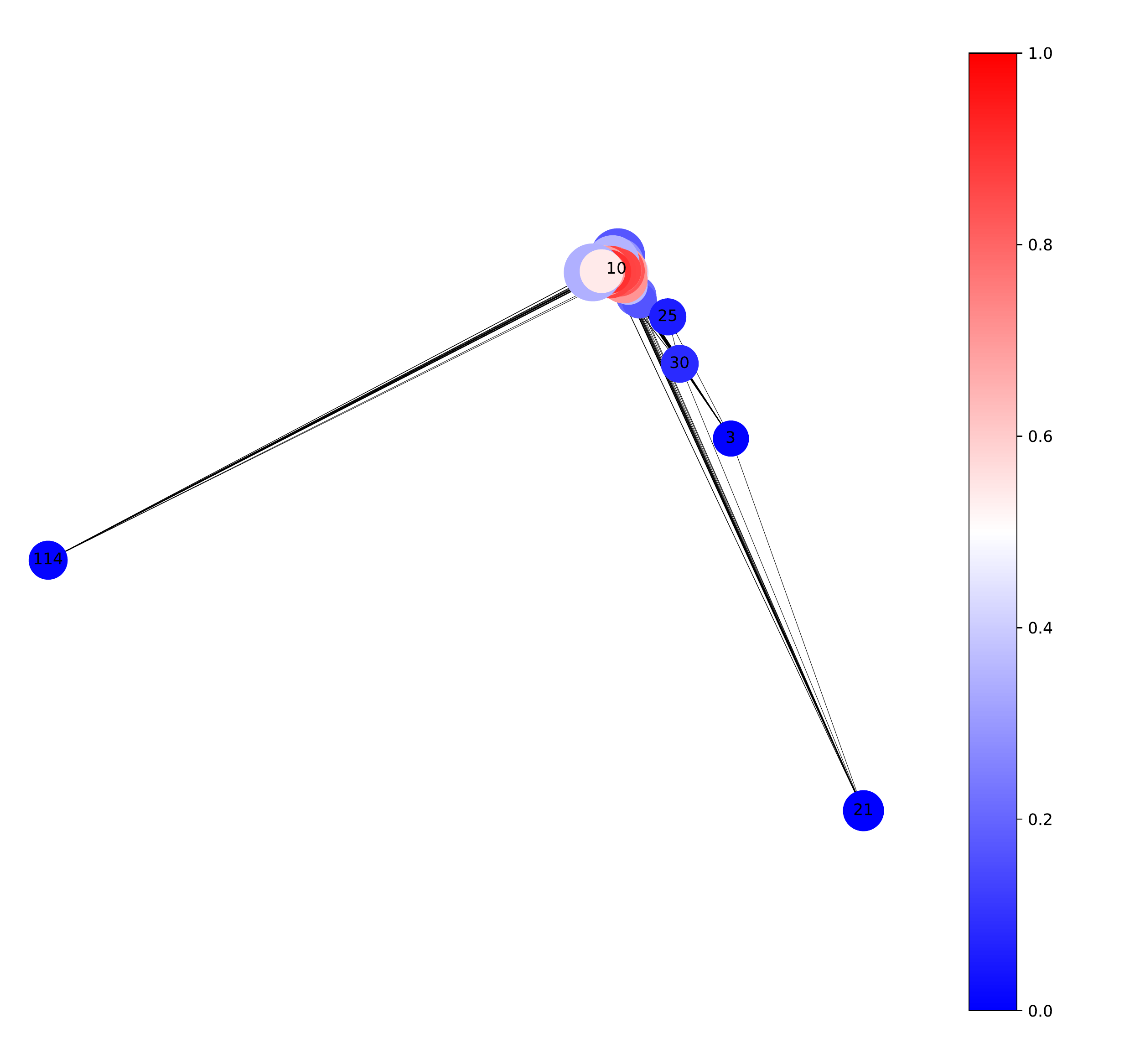}
    \caption{The graph of strain states in a spectral layout. The vertices are shaded by their eigenvector centralities, while the area of each vertex symbol is proportional to the Frobeius norm of the strain $\Vert\widehat{\boldsymbol{E}}_i\Vert_\text{F}$. The lower the eigenvector centrality of a vertex is, the longer are the edges connecting it to its neighboring vertices.}
    \label{fig:ms_spectral}
\end{figure}

The ability to visualize the distribution of the related, but simpler, degree centrality (or simply the degree of a vertex), over $\widetilde{G}$ reveals the strain states that are easy or difficult to attain. The Kamada-Kawai layout places the low degree centrality vertices at the periphery, and removed from the inner vertices that have high degree centrality, because their smaller numbers of edges allow a lower energy penalty in Equation \eqref{eq:graphenergy} even while the edges themselves are longer. This layout thus better delineates the lower degree vertices. Also of interest is the spectral layout \cite{Koren2005} in Figure \ref{fig:ms_spectral}, which is effective in accentuating the vertices with lower eigenvector centrality. This format shows the eigenvector components of each vertex for the second and third smallest eigenvalues of $\boldsymbol{L}$ on the Cartesian axes.\footnote{The minimum eigenvalue of $\boldsymbol{L}$ is zero.} Low eigenvector centrality vertices fall further out along these axes, helping to distinguish them from high eigenvector centrality vertices. Clearly, strain state $\widehat{\boldsymbol{E}}_{114}$ is related to the fewest states by this measure, followed by $\widehat{\boldsymbol{E}}_{21},\widehat{\boldsymbol{E}}_{3},\widehat{\boldsymbol{E}}_{30}$ and $\widehat{\boldsymbol{E}}_{25}$ in that order. For brevity we define $\widetilde{V}_\text{LEV} = \{\mathscr{S}_{114},\mathscr{S}_{21},\mathscr{S}_{3},\mathscr{S}_{30},\mathscr{S}_{25} \}$. The corresponding strain states for $\widetilde{\boldsymbol{E}}_{114}$ and $\widetilde{\boldsymbol{E}}_{21}$ are
\begin{alignat}{2}
    \widehat{\boldsymbol{E}}_{114} &=
    \begin{bmatrix}
     -0.015 & 0.017 & 0.002\\
     0.017 & -0.032 & 0.023\\
     0.002 & 0.023 & 0.003
    \end{bmatrix},\quad 
    \widehat{\boldsymbol{E}}_{21} &=
    \begin{bmatrix}
     -0.035 & 0.025 & -0.004\\
     0.025 & -0.010 & 0.013\\
     -0.004 & 0.013 & 0.025
    \end{bmatrix}
\end{alignat}

\subsubsection{Measure of states and eigenvector centrality of vertices}
\label{sec:elastcentrality}
Since, in Algorithm \ref{algo:strainstates}, the (non-)existence of edges is determined by (non-)convergence of the nonlinear solver, we investigated the strain states $\widehat{\boldsymbol{E}}_i$ of the three vertices $\mathscr{S}_i$ that lie furthest from the origin in the spectral layout. As a measure of the relative ``extremity'' of these states, we computed the Frobenius norms $\Vert\widehat{\boldsymbol{E}}_i\Vert_\text{F}$, which we represent by  linearly scaling the area of the vertices in the graph layouts. This measure has relevance because states with larger norms $\Vert\widehat{\boldsymbol{E}}\Vert_\text{F}$ lie further from the origin in $\mathbb{R}^6$, and also from low strain states. The strain state with maximum Frobenius norm (the most distant state from the origin in this norm in $\mathbb{R}^6$) is
\begin{equation}
    \widehat{\boldsymbol{E}}_{126} = 
    \begin{bmatrix}
    -0.038 & -0.013 & -0.095\\
    -0.013 & -0.004 & -0.047\\
    -0.095 & -0.047 & 0.025
    \end{bmatrix}
\end{equation}
and the maximum norm of the difference in states is $\Vert\widehat{\boldsymbol{E}}_{126} - \widehat{\boldsymbol{E}}_{86} \Vert_\text{F} = 0.1855$. From Figure \ref{fig:ms_kamadakawai}, it emerges that while $\mathscr{S}_{126}$ has low eigenvector centrality, it is not one of the five lowest: $\mathscr{S}_{126} \notin \widetilde{V}_\text{LEV}$. Thus, while $\widehat{\boldsymbol{E}}_{126}$ is a distant state by two measures in $\mathbb{R}^6$, it is far from being the least visited. We conclude that the magnitude of strain or of strain difference is not an indicator of states that have fewer reversible strain steps to other states. Instead, the interaction of the strain state with the underlying microstructure may play a dominant role in determining its accessibility.

\subsubsection{Cliques indicate sets of mutually accessible strain states}
\label{sec:elastcliquecycle}
The observations that there is a distribution of eigenvector centrality and degree centrality motivates investigation of other measures of connectedness between the strain states. The existence of cliques and cycles between states becomes relevant in this context. Using the Kamada-Kawai layout for its visual clarity, Figure \ref{fig:ms_clique} highlights the vertices belonging to the largest clique in red. Recall that this set of strain states are all reversibly attainable from each other. Of course, vertices belonging to a clique, say $\widetilde{G}_\alpha$, display a smaller maximum of the Frobenius norm of the difference between strain states, when compared with the maximum computed over the entire graph $\widetilde{G}$,
\begin{equation}
    \max\limits_{\mathscr{S}_i,\mathscr{S}_j\in\widetilde{G}_\alpha} \Vert \widehat{\boldsymbol{E}}_i - \widehat{\boldsymbol{E}}_j \Vert_\text{F} \le \max\limits_{\mathscr{S}_k,\mathscr{S}_l\in\widetilde{G}} \Vert \widehat{\boldsymbol{E}}_k - \widehat{\boldsymbol{E}}_l \Vert_\text{F}
\end{equation}
where the sub-graph over which the maximum on the left hand-side of the inequality is computed is any clique $\widetilde{G}_\alpha \subset \widetilde{G}$.

\subsubsection{Cycles}
\label{sec:elastcycles}
\begin{figure}
    \centering
    \includegraphics[scale=0.4]{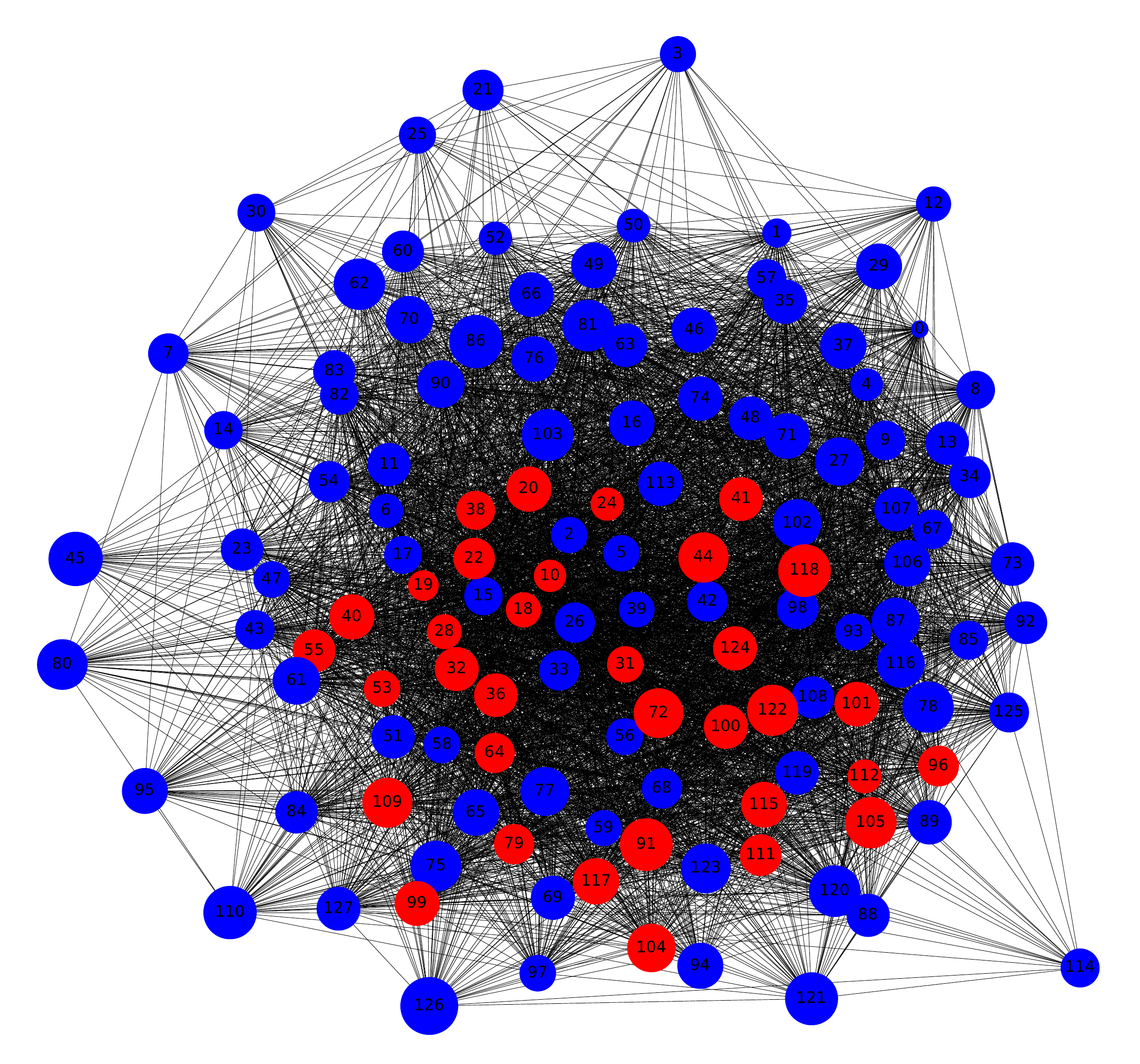}
    \caption{The largest clique in the graph of strain states is shown in red.}
    \label{fig:ms_clique}
\end{figure}

The number of cycles can be of interest, especially in sparsely connected graphs, where they are likely to be repeatedly traversed in the course of many walks. For the graph of strain states, cycles identify sequences of deformation that return to the starting state without reversing path, and suggest a measure of elastic cyclability of the microstructure. The graph under consideration harbors a large clique seen in Figure \ref{fig:ms_clique}, which as we now demonstrate, contains a very large number of cycles. The number of cycles in an undirected graph scales exponentially with $\vert\widetilde{V}\vert$: Consider the clique in Figure \ref{fig:ms_clique} as a subgraph $\widetilde{G}_\beta\subset\widetilde{G}$. It has $\vert\widetilde{V}_\beta\vert = 34$ and $\vert\widetilde{E}_\beta \vert= \binom{\vert\widetilde{V}_\beta\vert}{2} = 561$. Since $\widetilde{G}_\beta$ is undirected, it is the union of $\sum_{p = 1}^{561}2^p$ directed graphs.  Using the Rocha-Thatte algorithm \cite{Rocha2015} the number of cycles in each of these directed graphs can be found  in $p\sim 1,\dots, \binom{34}{2}$ iterations (the size of the longest path in each corresponding graph). Thus, the number of iterations needed to find all the cycles in $\widetilde{G}_\beta$ scales as $\sim \sum_{p = 1}^{561}p\cdot 2^p$. The graph $\widetilde{G} \supset \widetilde{G}_\beta$ itself will have many more cycles, which we have not attempted to enumerate.

\subsubsection{Shortest paths}
\label{sec:elastshortestpath}

\begin{figure}
    \centering
    \includegraphics[scale=0.4]{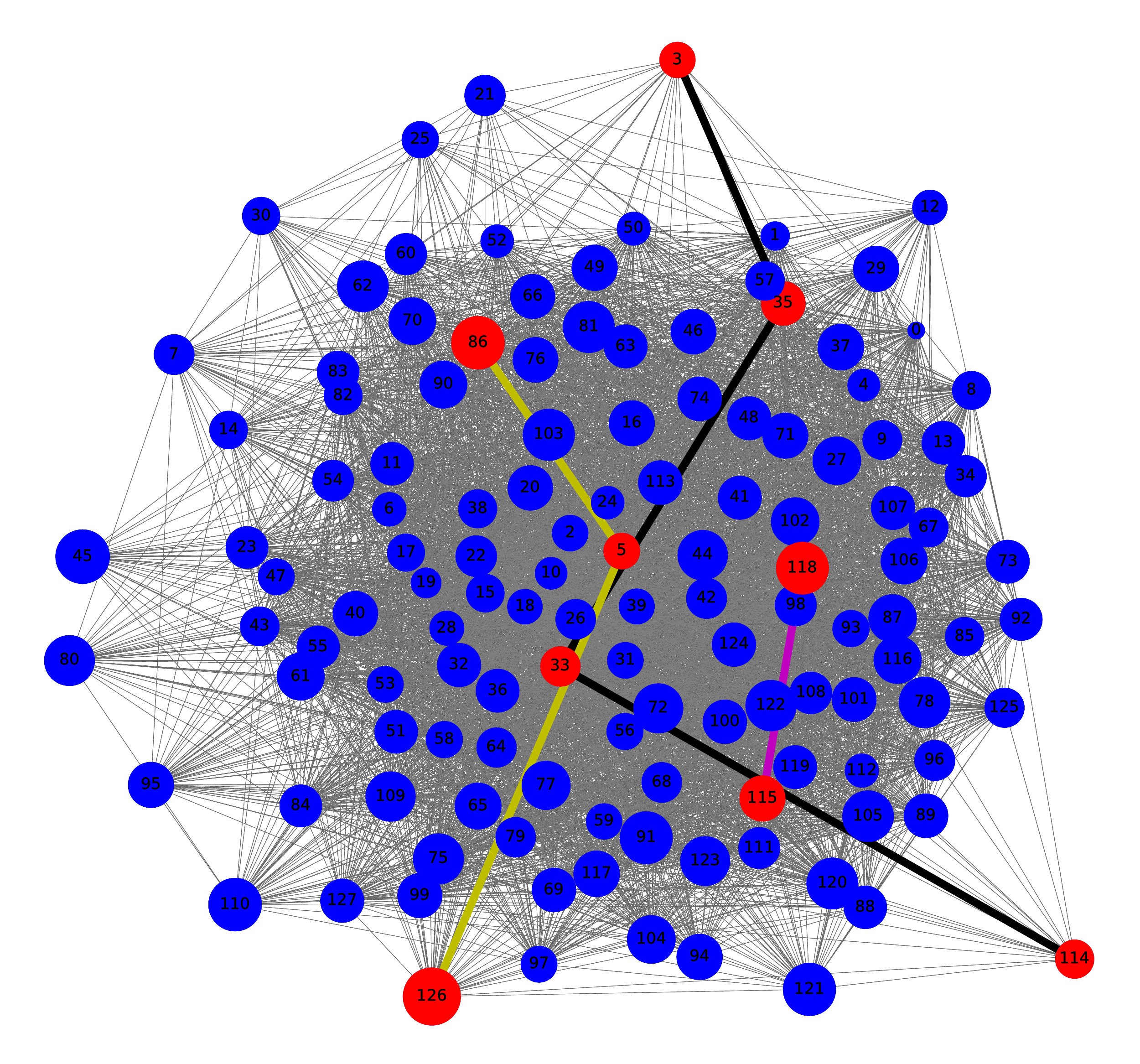}
    \caption{The shortest paths between the vertices in the pairs $\{\mathscr{S}_{86},\mathscr{S}_{126} \}$ (largest separation in $\widetilde{G}$), $\{\mathscr{S}_{115},\mathscr{S}_{118} \}$ (largest separation in the largest clique $\widetilde{G}_\beta \subset \widetilde{G}$) and $\{\mathscr{S}_{3},\mathscr{S}_{114} \}$ (largest separation in the Kamada-Kawai layout of $\widetilde{G}$).}
    \label{fig:ms_shortestpath}
\end{figure}

The final aspect of the graph we have studied is the shortest path between vertex pairs whose strain states are maximally separated over $\widetilde{G}$. Let $\{\mathscr{S}_i, \mathscr{S}_j\}$ be the unordered pair satisfying
\begin{equation}
    \{\mathscr{S}_i,\mathscr{S}_j \} = \text{arg}\,\max\limits_{\mathscr{S}_k,\mathscr{S}_l\in\widetilde{V}} \Vert \widehat{\boldsymbol{E}}_k - \widehat{\boldsymbol{E}}_l \Vert_\text{F}
    \label{eq:maxstraindiff}
\end{equation}
We find the shortest path between $\mathscr{S}_i, \mathscr{S}_j$. That is, if $\widetilde{V}_1 = \{\mathscr{S}_i,\mathscr{S}_k,\dots \mathscr{S}_l,\mathscr{S}_j \}$, $\widetilde{V}_2 = \{\mathscr{S}_i,\mathscr{S}_m,\dots \mathscr{S}_n,\mathscr{S}_j \}$, \dots, $\widetilde{V}_O = \{\mathscr{S}_i,\mathscr{S}_o,\dots \mathscr{S}_p,\mathscr{S}_j \}$ are distinct sets whose vertices are arranged in a sequence to define paths between $\mathscr{S}_i$ and $\mathscr{S}_j$, with corresponding edge sets $\widetilde{E}_1,\widetilde{E}_2,\dots \widetilde{E}_O$, we seek
\begin{equation}
    \widetilde{E}_S = \text{arg} \min\limits_{\widetilde{E}_1,\dots \widetilde{E}_O} \vert \widetilde{E}_K\vert
\end{equation}
Thus, $\widetilde{E}_S$ gives us the most efficient path through the most distantly separated strain states of those sampled to generate data for numerical homogenization of the microstructure. For the strain states graph, $\widetilde{G}$, the pair corresponding to Equation \eqref{eq:maxstraindiff} is $\{\mathscr{S}_{86}, \mathscr{S}_{126}\}$. Figure \ref{fig:ms_shortestpath} shows the shortest path between the vertices corresponding to this pair, as well as for $\{ \mathscr{S}_{115},\mathscr{S}_{118}\}$, which is the corresponding pair for the clique $\widetilde{G}_\beta$. For comparison, we also have shown the shortest path between  $\{\mathscr{S}_3, \mathscr{S}_{114}\}$, which is the most separated pair in the Kamada-Kawai layout.

\subsubsection{Components}
\label{sec:elastcomponents}
We note that the rapidly fluctuating strains $\boldsymbol{E}$ over each of the martensitic microstructures in Figure \ref{fig:microstructures} are associated with a finely corrugated energy surface in the high-dimensional space in which we have obtained these numerical solutions.\footnote{This finely corrugated energy surface arises from the non-convex component, $\widetilde{\psi}$ of the free energy density in Equation \eqref{eq:freeenergy}.} As a consequence, while, physically, there exist trajectories in strain space $\boldsymbol{E}(\boldsymbol{\xi}) \in \mathbb{R}^6$, with a non-monotonic dependence on a vector parameter $\boldsymbol{\xi}$, that transform one microstructure of Figure \ref{fig:microstructures} into another,  it is very challenging to numerically trace such transitions. Even if trajectories were identified, numerical solutions along them could prove stiff to the point of non-convergence. For this reason, the graph associated with each microstructure in Figure \ref{fig:microstructures}, say $\widetilde{G}_{\text{ms}_1}, \widetilde{G}_{\text{ms}_2}, \dots$, forms a connected component of a larger graph $G = \widetilde{G}_{\text{ms}_1}\cup \widetilde{G}_{\text{ms}_2}\cup\dots$, which is itself not fully connected. The problem of numerically finding those edges that would make $G$ fully connected is of mainly mathematical, rather than practical interest because of the difficulty of following such trajectories.


    \label{fig:stat-nonlin-graph}

\subsection{Graphs constructed on time series solutions of a dissipative dynamical system }
\label{sec:CHdynamicsgraphs}

Thermodynamic dissipation is near-ubiquitous in dynamic physical processes. Here, we consider a first-order, dissipative, dynamical problem related to phase transformations in materials physics and biophysics, and seek to study it in the context of the properties observed in the abstract in Section \ref{sec:diss-dynamics}.

\subsubsection{First-order dynamics of a two-species, phase-separating system}
\label{sec:CHsystem}
\begin{figure}
    \centering
    \includegraphics[scale=0.3]{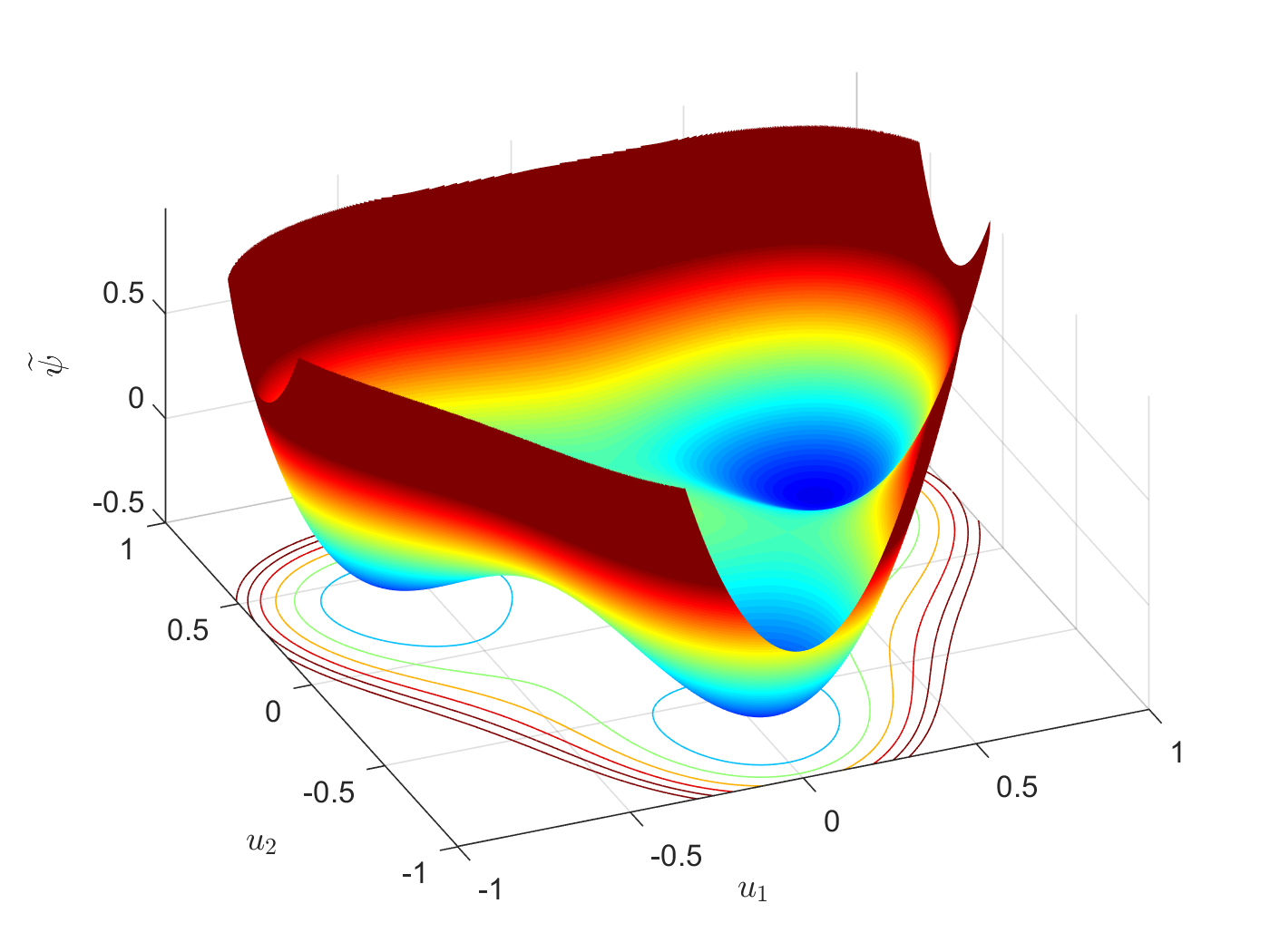}
    \caption{The non-convex free energy density function, $\widetilde{\psi}(u_1,u_2)$ that gives rise to separation into three phases corresponding to the three wells.}
    \label{fig:freeenergy2}
\end{figure}

We base our treatment on the Cahn-Hilliard equation written for two species whose compositions $u_1$ and $u_2$ satisfy $-1 \le u_1,u_2 \le 1$. The Cahn-Hilliard equation is a conservative phase field method that models phase separation as the consequence of an instability in a nearly uniform composition field. The description is based on a free energy density function of the form
\begin{equation}
    \psi(u_1,u_2,\nabla u_1,\nabla u_2) = \widetilde{\psi}(u_1,u_2) + \frac{1}{2}\kappa_1\Vert \nabla u_1\Vert^2 + \frac{1}{2}\kappa_2\Vert \nabla u_2\Vert^2,
    \label{eq:CHfreeenergy}
\end{equation}
where $\widetilde{\psi}(u_1,u_2)$ is non-convex with respect to $u_1$ and $u_2$, and the gradient terms with coefficients $\kappa_1$ and $\kappa_2$ represent interfacial energies. Of interest here is a function of the form 
\begin{equation}
\widetilde{\psi}(u_1,u_2) =  \frac{3 d}{2 s^4} (u_1^2+u_2^2)^2 + \frac{d}{s^3} u_2 (u_2^2-3u_1^2) - \frac{3 d}{2 s^2} (u_1^2+u_2^2), \quad d,s > 0
\label{eq:homogenergy3well2field}
\end{equation}
This ``homogeneous'' component of the free energy density is illustrated in Figure \ref{fig:freeenergy2}. The three local minima represent distinct phases that are explained below.

Chemical potentials are defined via variational derivatives of the total free energy density:
\begin{subequations}

\begin{alignat}{2}
    \mu_1 &= \frac{\delta \psi}{\delta u_1} &&= \frac{\partial\widetilde{\psi}}{\partial u_1} - \kappa_1\nabla^2 u_1
    \label{eq:chempot1}\\
    \mu_2 &= \frac{\delta \psi}{\delta u_2} &&= \frac{\partial\widetilde{\psi}}{\partial u_2} - \kappa_2\nabla^2 u_2
    \label{eq:chempot2}
\end{alignat}
\end{subequations}
The governing equations are first-order in time and of conservation form
\begin{subequations}
\begin{alignat}{3}
    \frac{\partial u_i}{\partial t} &= M_i\nabla^2\mu_i  &&\text{in} &&\Omega\times[0,T]\\
    &= M_i(\nabla^2\frac{\partial\widetilde{\psi}}{\partial u_i} - \kappa_i \nabla^4 u_i) &&\text{in} &&\Omega\times[0,T]\label{eq:CHgoveq}\\
    \nabla\left(\frac{\partial\widetilde{\psi}}{\partial u_i} - \kappa_i\nabla^2 u_i\right)\cdot\boldsymbol{n} &=0 &&\text{on} &&\partial\Omega\times[0,T]\label{eq:CHzeroflux}\\
    \nabla u_i\cdot\boldsymbol{n} &= 0 &&\text{on} &&\partial\Omega\times[0,T]\label{eq:CHhodirichlet}
\end{alignat}
\end{subequations}

\noindent for $i = 1,2$. The surface normal is $\boldsymbol{n}$. The coefficients defining $\widetilde{\psi}$ in Equations \eqref{eq:CHfreeenergy} and \eqref{eq:homogenergy3well2field}, and the mobilities, $M_1, M_2$ in the governing equations \eqref{eq:CHgoveq} appear in Table \ref{tbl:ch3well2field}.
\begin{table}[h]
\centering
\caption{Parameters for the two-species phase-separation problem.}
\begin{tabular}{ |c|c|c|c|c|c|c|  }
\hline
 Parameter & $d$ & $s$ & $\kappa_1$ & $\kappa_2$ & $M_1$ & $M_2$ \\
 \hline
 Value & $0.4$ & $0.7$  & $1$ or $10$ & $1$ & $0.1$ & $0.1$\\
 \hline
\end{tabular}
 \label{tbl:ch3well2field}
\end{table}

\subsubsection{Graphs constructed on dynamic states indexed by time}
\label{sec:CHgraphs}

\begin{figure}
    \centering
    \includegraphics[scale=1.0]{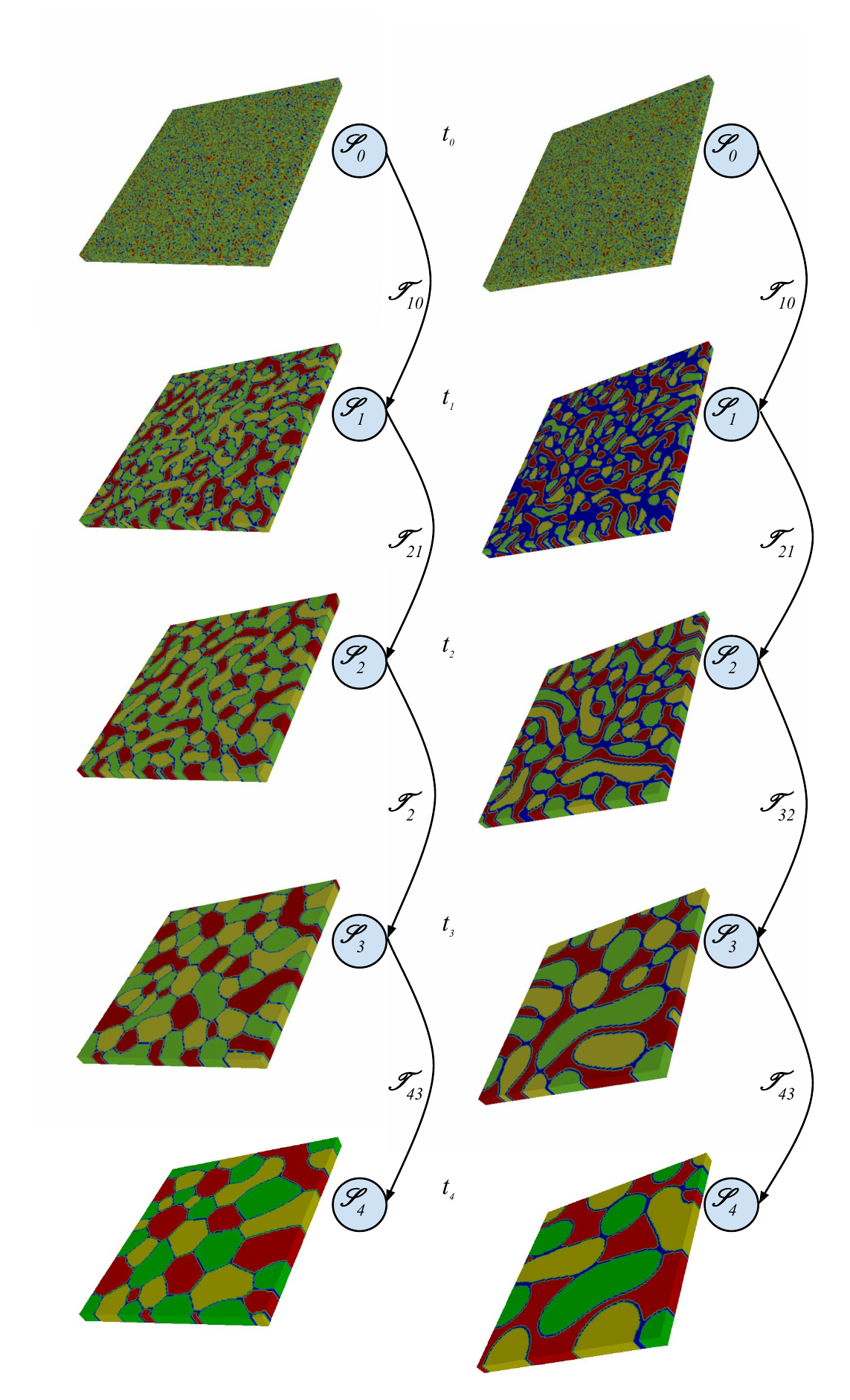}
    \caption{Graphs, $G_1$ (left) and $G_2$ (right) representing computations on two IBVPs of the Cahn-Hilliard equation with distinct, randomized initial conditions and differing gradient (interface) energy parameters. Separation is seen into three phases, but corresponding states in each graph are different due to the distinct initial conditions and gradient energy parameters.}
    \label{fig:dissip2-comp}
\end{figure}

Figure \ref{fig:dissip2-comp} shows graphs, $G_1$ on the left and $G_2$ on the right, which represent distinct IBVPs modelled by Equations (\ref{eq:CHfreeenergy}--\ref{eq:CHhodirichlet}). A state of the system can be defined as the vector $\mathscr{S}_i = (\theta_{1_i}, \theta_{2_i}, \theta_{3_i}, \Pi_i, \Gamma_i, \Gamma_{12_i})$, corresponding to time $t = t_i$. Here, $\theta_{1_i},\theta_{2_i},\theta_{3_i}$ are the volumes of the phases corresponding to the minima at $(0.35,0.61), (-0.35,0.61)$ and $(0,-0.7)$, respectively, in the $u_1-u_2$ plane, $\Pi_i = \int_\Omega \psi_i\mathrm{d}V$ is the total free energy of the state, $\Gamma_i$ is the total interfacial energy of the state and $\Gamma_{12_i}$ is the interfacial energy between the $\theta_{1}$ and $\theta_2$ phases in that state. The edges, $\mathscr{T}_{ij}$ are defined by the nonlinear, time-stepping solution scheme as proposed in Section \ref{sec:diss-dynamics-properties}. In Figure \ref{fig:dissip2-comp} the phases with volumes $\theta_{1},\theta_2$ and $\theta_3$ appear in green, yellow and red, respectively. The initial conditions in states $\mathscr{S}_0$ are randomized for each graph and appear as they were computed. The dissipation in this problem is manifested in a decreasing free energy density, $\Pi$ as the system progresses through time-parameterized states as illustrated in Figure \ref{fig:dissip2-comp}. Since $\dot{\Pi}\le 0$, $\Pi$ itself is an entropy quantity, and renders the graphs directed following the discussion of properties of such systems in Section \ref{sec:diss-dynamics-properties}. Since the free energy, $\Pi_i$, of a state, $\mathscr{S}_i$, depends on the path taken to reach the state, it follows that a path, starting at $\mathscr{S}_i$, but not following exactly the sub-path $\mathscr{S}_i\rightarrow\mathscr{S}_j\rightarrow\mathscr{S}_k\rightarrow\mathscr{S}_l$ already present in the graph, cannot arrive at $\mathscr{S}_l$. At least the value of $\Pi$ will differ. The five states shown in Figure \ref{fig:dissip2-comp} for each graph at $t = t_0,t_1,t_2,t_3,t_4$ are only for illustrative purposes. Our computations yielded many more states: $\vert V_1\vert= 1125$ and $\vert V_2\vert = 1252$.

Because the graphs $G_1$ and $G_2$ were generated from the system of partial differential equations (\ref{eq:CHfreeenergy}--\ref{eq:CHhodirichlet}), which impose a sequence of states by their dissipative character, these directed graphs are linear, unbranched trees. Their unweighted adjacency matrix entries are 
\begin{equation}
    A_{ij} =\begin{cases}
                 1 & \text{if}\; j = i-1\\
                 0 & \text{otherwise}
            \end{cases}
\end{equation}
yielding $\boldsymbol{A}_1$ and $\boldsymbol{A}_2$ that are lower sub-diagonal, unlike the symmetric adjacency matrix for the strain states graph (see Section \ref{sec:elastgraphlayouts}). Because of their simplicity, there is not much insight to be gained by analyzing these graphs. However, having constructed the graphs by extracting the relatively low-dimensional state vectors $\mathscr{S}_i \in \mathbb{R}^6$ from the original finite element computations on $\sim 10^6$ degrees of freedom (appearing in Figure \ref{fig:dissip2-comp}), we are afforded other approaches to study the system. 

\subsubsection{Dynamics of low-dimensional states}
\label{sec:CHevolstates}
The differences that can be seen between corresponding states, say $\mathscr{S}_2$, on $G_1$ and $G_2$ in Figure \ref{fig:dissip2-comp} arise from different values of $\kappa_1$ used in Equation \eqref{eq:CHgoveq} to generate the respective graphs. Recall that the coefficients $\kappa_1$ and $\kappa_2$ impose penalties on gradients of $u_1$ and $u_2$,  respectively, thereby introducing interfacial energy densities in Equation \eqref{eq:CHfreeenergy}. By increasing $\kappa_1$ by an order of magnitude (from $1$ to $10$), interfaces between the phases with volumes $\theta_1$ and $\theta_2$ (green and yellow) are made more sharply unfavorable in the computations that generate $G_2$. This leads to the shorter total interface length between these phases, which is apparent on comparing the states  $\mathscr{S}_2,\mathscr{S}_3$ and $\mathscr{S}_4$ across graphs $G_1$ and $G_2$ in Figure \ref{fig:dissip2-comp}. It is instructive to study the divergence of the system dynamics in these two cases via the state quantities $\mathscr{S}_i = (\theta_{1_i}, \theta_{2_i}, \theta_{3_i}, \Pi_i, \Gamma_i, \Gamma_{12_i})$ represented in the graph vertices. 

\begin{figure}
    \centering
    \includegraphics[scale=0.22]{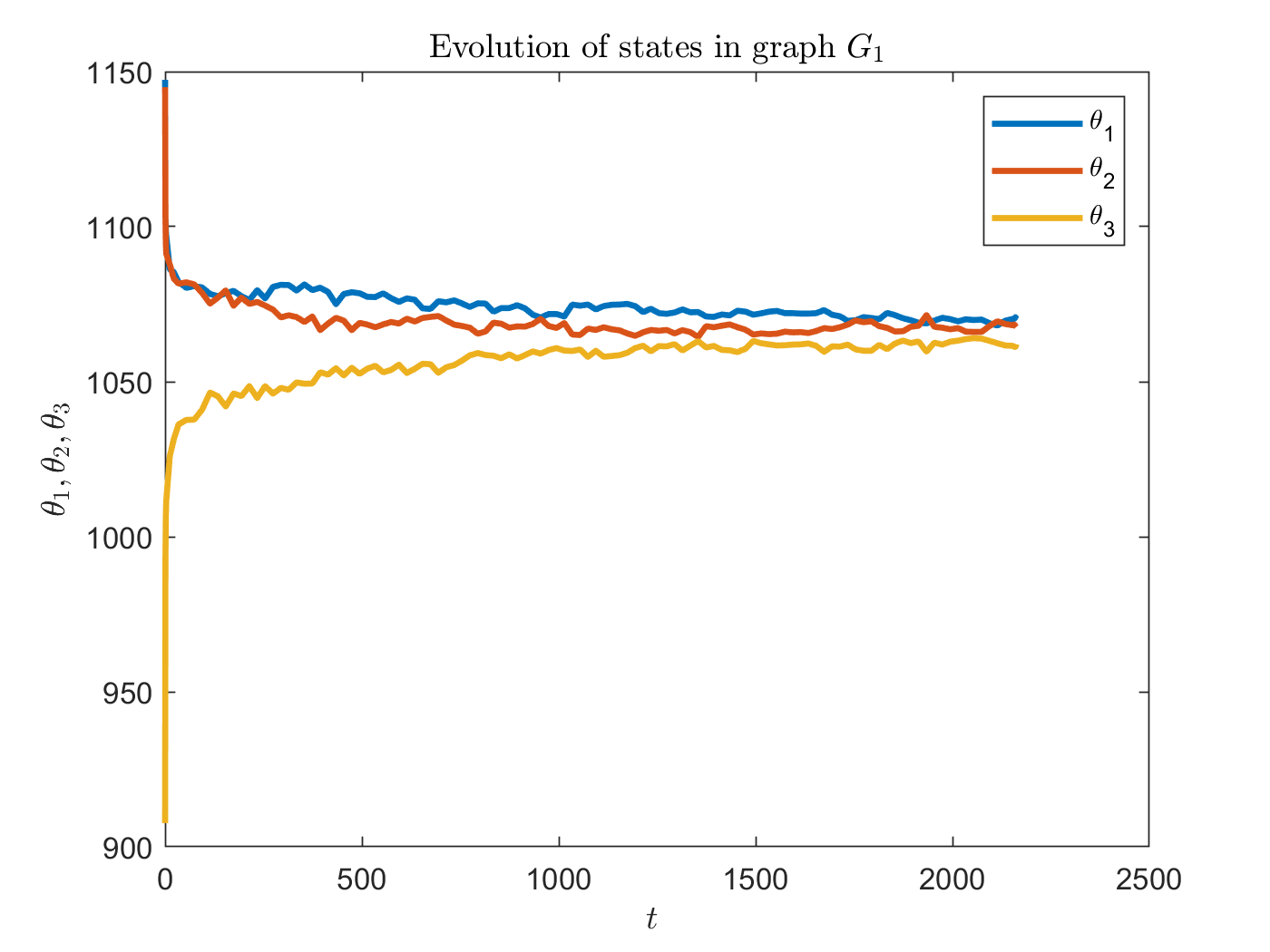}
    \includegraphics[scale=0.22]{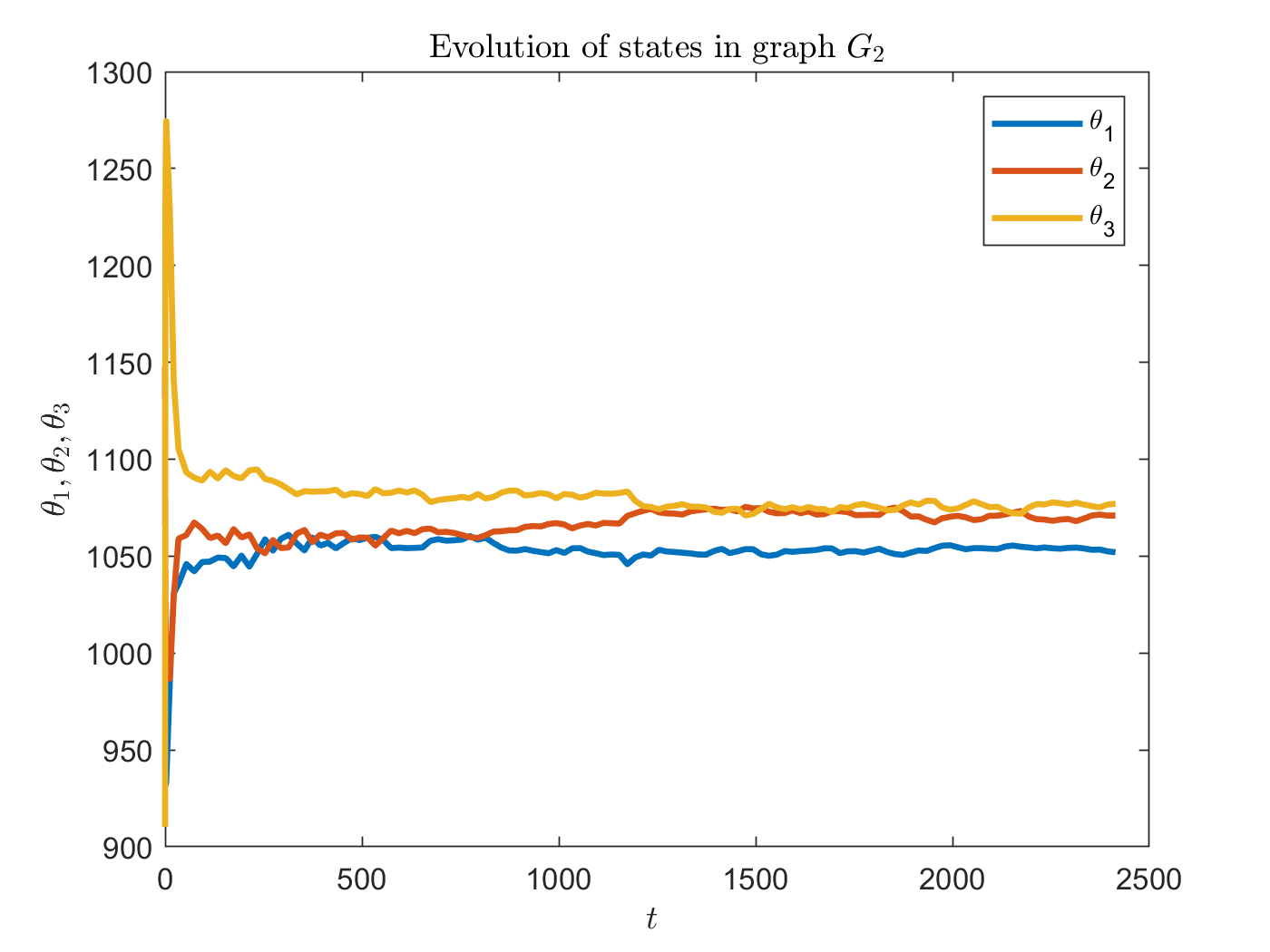}
    \caption{Evolution of phases $\theta_1,\theta_2,\theta_3$ for graphs $G_1$ (left) and $G_2$ (right).}
    \label{fig:theta}
\end{figure}

Figure \ref{fig:theta} shows the volumes $\theta_1,\theta_2$ and $\theta_3$ evolving with time.\footnote{The dynamics have been shown with respect to time rather than states in order to allow a physically meaningful comparison between the graphs. Because different time steps were used in the computations, states indexed by the same number in $G_1$ and $G_2$ do not necessarily correspond to the same time in the dynamics, unlike the schematic illustration in Figure \ref{fig:dissip2-comp}.} At the final time of $t = 2162$ for the IBVP represented by $G_1$, these quantities are $\theta_1 = 1070, \theta_2 = 1068$ and $\theta_3 =  1062$ for a mean $\theta_\text{m} = 1066.67$. At the same time for the longer running computation represented by $G_2$, the corresponding values are $\theta_1 = 1055, \theta_2 = 1069$ and $\theta_3 =  1077$ with a mean $\theta_\text{m} = 1067$. The deviation of the curve of $\theta_1$ from $\theta_2$ and $\theta_3$ for $G_2$ appears to be driven by the specific instance of randomized initial conditions, especially given that the higher $\kappa_1$ for the corresponding IBVP introduces a mechanism that does not favor $\theta_1$ or $\theta_2$ relative to each other. We conclude, therefore, that the phase volumes are not the ideal quantities for conveying insight to the macroscopic evolution of the system.

\begin{figure}
    \centering
    \includegraphics[scale=0.22]{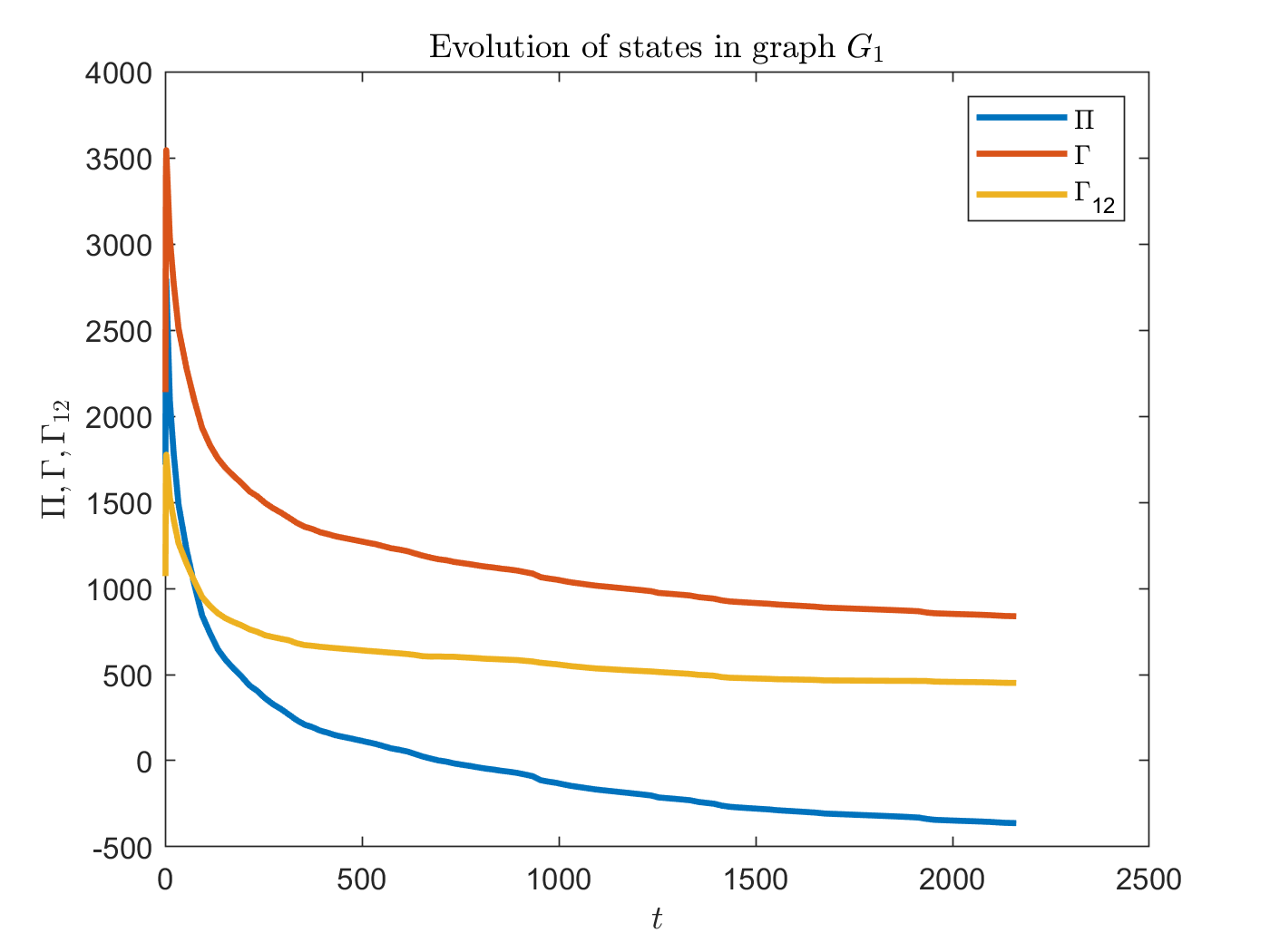}
    \includegraphics[scale=0.22]{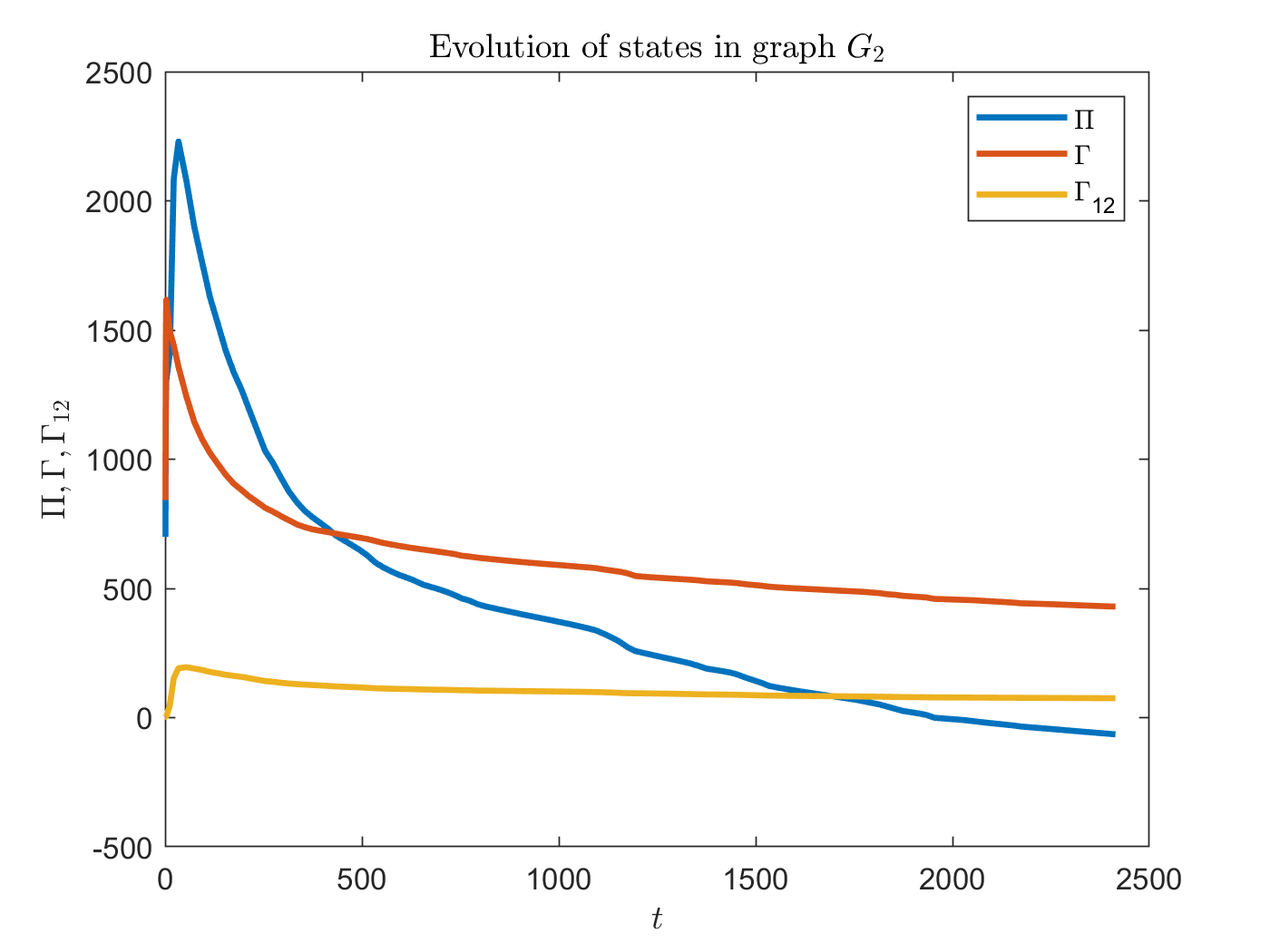}
    \caption{Evolution of total free energy, $\Pi$, total interfacial energy, $\Gamma$ and $\theta_1-\theta_2$ interfacial energy $\Gamma_{12}$ for graphs $G_1$ (left) and $G_2$ (right).}
    \label{fig:CHenergies}
\end{figure}

The free energies in Figure \ref{fig:CHenergies} offer more insight. We first note that the total free energy, $\Pi$, is decreasing for all except very early times.\footnote{The free energy $\Pi$ does show increases at very early times $t \le 2.9$ for $G_1$ and $t \le 20$ for $G_2$. This is a numerical artifact that is not germane to the graph theoretic methods being proposed in this communication. The backward Euler time integration scheme used for this nonlinear IBVP does not guarantee a non-increasing free energy. The very rapid dynamics of spinodal decomposition, which occurs at early times in this problem requires much smaller time steps to ensure non-increasing free energy than were taken for this computation.} The distinction between $G_1$ and $G_2$ is brought out by the positive definite interfacial free energies, especially $\Gamma_{12}$, which corresponds to the $\theta_1-\theta_2$ phase interface: At $t = 2162$ this quantity for $G_1$ is $\Gamma_{12} = 452$, and for $G_2$, it is $\Gamma_{12} = 77$. Related differences are seen in the total interfacial free energy, $\Gamma$, which includes a contribution from $\Gamma_{12}$, The physical relevance is clear: the higher penalty imposed on $\theta_1-\theta_2$ interfaces by $\kappa_1 = 10$ decreases their total length in the computations represented in $G_2$ (see the green-yellow interfaces in Figure \ref{fig:dissip2-comp}) leading to a significantly lower interfacial energy.

\subsubsection{Dissipative dynamics as an organizing principle for graph generation}
\label{sec:CHorgprinciples}
It is noteworthy that, while the physics of the first-order dissipative dynamical system in Equations (\ref{eq:CHfreeenergy}--\ref{eq:CHhodirichlet}) may be considered more complex than the stationary system in Equations (\ref{eq:defgrad}-\ref{eq:freeenergy}) by some measures, the complexity of the resulting graphs is reversed. The non-branching, time-series dynamics and thermodynamic dissipation jointly impose an organizing principle, so that the graphs representing the IBVPs of the Cahn-Hilliard equation are linear, directed trees. An organizing principle, however, needs to be imposed on the states of stationary systems. By identifying the nonlinear solution scheme between strain states as an organizing principle that induces edges, the corresponding graphs are imbued with rich structure as elucidated in Section \ref{sec:nonconvexelasticity}. The graphs induced by first-order dynamics could have greater complexity if, instead of a single sequence of states dictated by the temporal evolution of the dynamics, different solution paths were explored over time as branches from some chosen states. The graphs $G_1$ and $G_2$, however, would remain trees; cliques and cycles would still be absent.

\subsection{Graphs constructed on the states of a dissipative dynamical system without time series data}
\label{sec:DNS-MLgraphs}

We continue to explore the role of thermodynamic dissipation as an organizing principle for graph generation. However, we consider states that have not been indexed by time, unlike the case in Section \ref{sec:CHdynamicsgraphs}. As we demonstrate, the second law of thermodynamics leads to a transition quantity that induces edges between states to construct graphs. When considered in light of the properties identified in Section \ref{sec:diss-dynamics}, the corresponding graphs are richer than those in Section \ref{sec:CHdynamicsgraphs}.

\subsubsection{First-order dynamics of a phase transforming binary alloy system}

\begin{figure}[tb]
        \centering
        \includegraphics[width=0.55\textwidth]{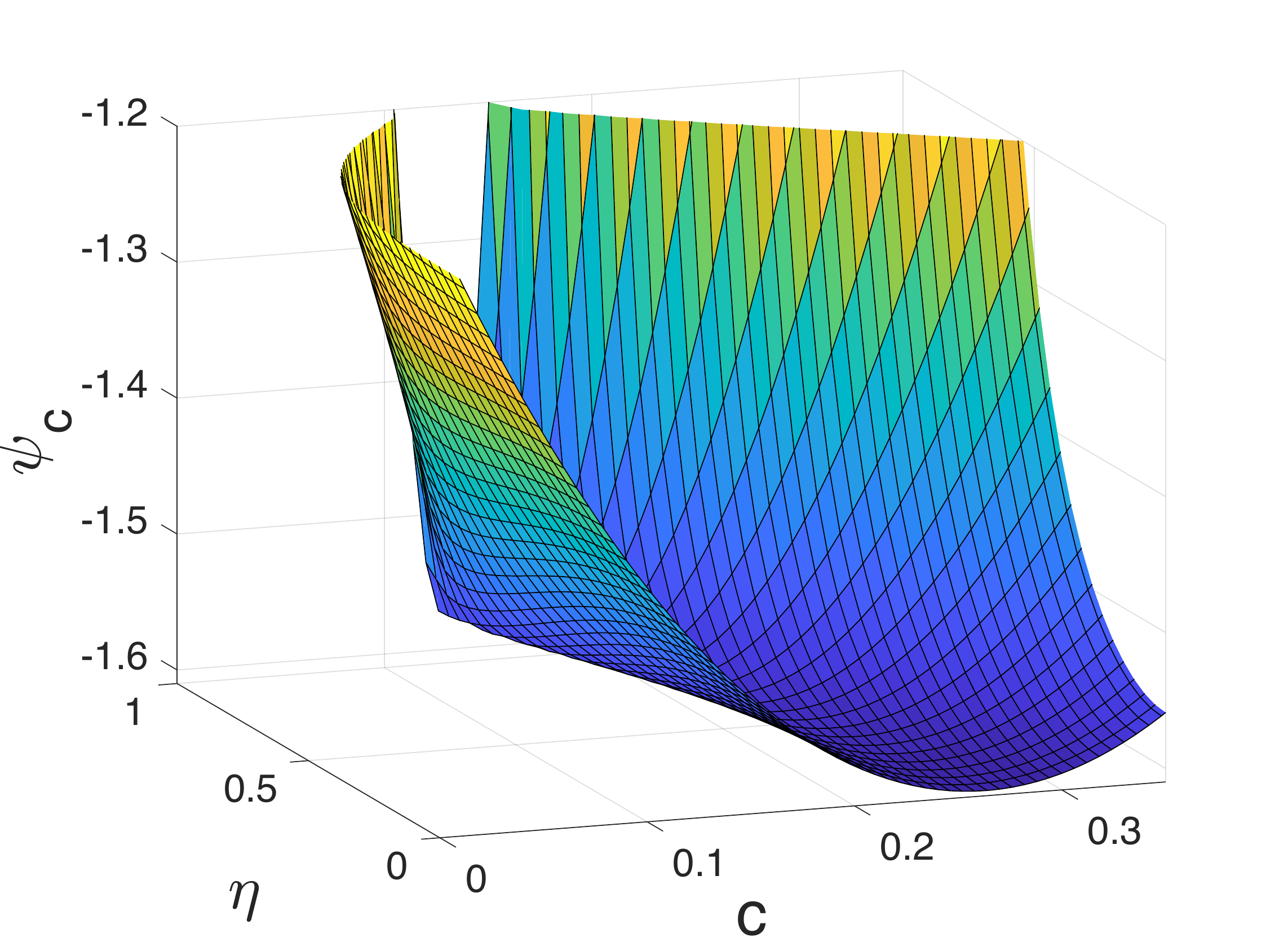}
        	\caption{Plot of the local chemical free energy density, $\psi_\text{c}$ \eqref{eq:chemfreeenergy}, in the Mg-Y alloy for transformations between $\alpha$ (Mg) and $\beta^\prime$ (Mg-Y) phases. The wells at $\eta = 0$ and $\eta = 1$ correspond to the $\alpha$ and $\beta^\prime$ phases.}
	\label{fig:localFreeEnergy}
\end{figure}

Phase field methods such as the Cahn-Hilliard equation, whose states generated the graphs of Section \ref{sec:CHdynamicsgraphs}, are characterized by their imposition of first-order dynamics to traverse a free energy landscape. The dynamics of phase field methods are subject to the second law's requirement of non-increasing free energy. Another variant of phase field methods is the Allen-Cahn equation \cite{AllenCahn1979}, which differs from the Cahn-Hilliard equation in having non-conservative dynamics, and representing phase transformations by nucleation and growth of precipitates. It too has widely been applied to model precipitate formation during phase transformations in binary alloy systems. The basis of this treatment also lies in a free energy density, now defined as a function of composition, $c$, an order parameter, $\eta$, its gradient $\nabla\eta$ and the elastic deformation gradient, $\boldsymbol{F}^\text{e}$, \cite{Kim1999,jietal2014}
\begin{equation}
\psi(c,\eta,\nabla\eta,\boldsymbol{F}^\text{e}) = \psi_\text{c}(c,\eta) + \psi_\text{grad}(\nabla\eta) + \psi_\text{e}(\boldsymbol{F}^\text{e}(\eta,\boldsymbol{F}),\eta)
\label{eq:Pi}
\end{equation}
where $\psi_\text{c}(c,\eta)$ is the local chemical free energy density, $\psi_\text{grad}(c,\eta)$ is the gradient energy term, and $\psi_\text{e}(\boldsymbol{F}^\text{e}(\eta,\boldsymbol{F}),\eta)$ is the elastic strain energy density. The deformation gradient, $\boldsymbol{F}$, reappears, and its elastic component is $\boldsymbol{F}^\text{e}$, which is defined below. We use
\begin{align}
\psi_\text{c}(c,\eta) &= \psi_\text{c}^\alpha(c^\alpha)\left(1-h(\eta)\right)
+\psi_\text{c}^{\beta^\prime}(c^{\beta^\prime})h(\eta)
+\omega \widetilde{\psi}(\eta)\label{eq:chemfreeenergy}\\
h(\eta) &= 3\eta^2 - 2\eta^3\\
\widetilde{\psi}(\eta) &= \eta^2-2\eta^3+\eta^4 \label{eqn:Landau}
\end{align}
Here, $\psi_\text{c}$ is written for a Mg-Y alloy in terms of contributions from the matrix phase, $\alpha$ (Mg) and precipitate phase, $\beta^\prime$ (Mg-Y). The regularized Heaviside function $h(\eta)$ interpolates smoothly between the phases, using $h(0) = 0$, $h(1) = 1$, and $h'(0) = h'(1) = 0$. As seen from Equation \eqref{eq:chemfreeenergy} the $\alpha$ phase corresponds to $\eta = 0$, and the $\beta^\prime$ phase to $\eta = 1$. The Landau free energy density, $\widetilde{\psi}$, drives the  structural change in the alloy, and has wells for the $\alpha$ and $\beta^\prime$ phases, respectively, at $\eta = 0$ and $\eta = 1$. The functions $\psi_\text{c}^\alpha(c^\alpha)$ and $\psi_\text{c}^{\beta^\prime}(c^{\beta^\prime})$ are written as quadratic approximations:
\begin{align}
    \psi_\text{c}^\alpha(c^\alpha) &\approx A^\alpha(c^\alpha - c^\alpha_0)^2 + B^\alpha \label{eqn:f_alpha}\\
    \psi_\text{c}^{\beta^\prime}(c^{\beta^\prime}) &\approx A^{\beta^\prime}(c^{\beta^\prime} - c^{\beta^\prime}_0)^2 + B^{\beta^\prime} \label{eqn:f_beta}
\end{align}
where the parameters appear in Table \ref{tab:chem2}. The quantities $c^\alpha$ and $c^{\beta^\prime}$ are obtained from constraint conditions on chemical potentials and the alloy composition, and have the forms:
\begin{align}
c^\alpha &= \frac{A^{\beta^\prime}\left[c - c^{\beta^\prime}_0h(\eta)\right] + A^\alpha c^\alpha_0h(\eta)}{A^\alpha h(\eta) + A^{\beta^\prime}(1 - h(\eta))} \label{eqn:calpha}\\
c^{\beta^\prime} &= \frac{A^\alpha\left[c - c^\alpha_0(1-h(\eta))\right] + A^{\beta^\prime} c^{\beta^\prime}_0(1-h(\eta))}{A^\alpha h(\eta) + A^{\beta^\prime}(1 - h(\eta))}  \label{eqn:cbeta}
\end{align}
The resulting local free energy density, $\psi_\text{c}$, appears in Figure \ref{fig:localFreeEnergy}.

\begin{table}[tb]
    \centering
    \caption{Parameters in the quadratic chemical free energy density descriptions \eqref{eqn:f_alpha} and \eqref{eqn:f_beta}.}
    \begin{tabular}{c | c c}
    \hline
    $A^\alpha$ & 6.2999 &GPa\\
    $B^\alpha$ & -1.6062 &GPa\\
    $c^\alpha_0$ & 0.2635\\
    $A^{\beta^\prime}$ & 704.23 &GPa\\
    $B^{\beta^\prime}$ & -1.5725 &GPa\\
    $c^{\beta^\prime}_0$ & 0.1273
    \end{tabular}
    \label{tab:chem2}
\end{table}

The gradient energy term is defined via a second order tensor $\boldsymbol{\kappa}$
\begin{align}
\psi_\text{grad}(\nabla\eta) &= \frac{1}{2}\nabla\eta\cdot\boldsymbol{\kappa}\nabla\eta
\label{eqn:grad_energy}
\end{align}
 and is anisotropic if $\boldsymbol{\kappa}$ is an anisotropic tensor. The components of $\boldsymbol{\kappa}$ are related to the barrier height $\omega$, the interface thickness, and the anisotropic interfacial energies based on the equilibrium solution for the one-dimensional problem and neglecting elasticity. Details of these approximations appear in the work of Kim et al. \cite{Kim1999} and Teichert \& Garikipati \cite{Teichert2018}, while the interfacial energies themselves were computed and reported by Liu et al. \cite{Liu2013}. Following Teichert \& Garikipati \cite{Teichert2018} we use
\begin{align}
\boldsymbol{\kappa} = 
\begin{bmatrix}
0.1413 & 0 & 0\\
0 & 0.002993 & 0\\
0 & 0 & 0.1197
\end{bmatrix},\;\omega = 0.115896
\end{align}

The strain energy density function is an anisotropic St.\ Venant-Kirchhoff model. The elasticity constants are modeled as dependent on the order parameter to represent the difference in elasticity between the two phases. The total strain energy of the precipitate-matrix system is driven by a strain mismatch between the crystal structures of the matrix, $\alpha$-, and precipitate, $\beta^\prime$-phases. The stress-free transformation tensor of the $\beta^\prime$ precipitate, $\boldsymbol{F}_{\beta^\prime}$ (see Table \ref{tab:eigenstrain}), and the order parameter determine the misfit eigenstrain, represented by $\boldsymbol{F}^\lambda$.
\begin{table}[tb]
    \centering
    \caption{Components of the deformation gradient tensor, representing the eigenstrain in the Mg-Y $\beta^\prime$ precipitate \cite{Natarajan2017}.}
    \begin{tabular}{c | c }
     & $c_{\mathrm{Y}} = 0.125$\\
    \hline
    $F_{\beta^\prime_{11}}$ & 1.0307\\
    $F_{\beta^\prime_{22}}$ & 1.0196\\
    $F_{\beta^\prime_{33}}$ & 0.9998
    \end{tabular}
    \label{tab:eigenstrain}
\end{table}
The multiplicative decomposition of the total deformation gradient into elastic and the misfit components is also defined:
\begin{subequations}
\begin{align}
\psi_\text{e}(\boldsymbol{F}^\text{e}(\eta,\boldsymbol{F}),\eta) &= \frac{1}{2}\boldsymbol{E}^\text{e}:(\mathbb{C}^\alpha(1-h(\eta) + \mathbb{C}^{\beta^\prime}h(\eta)):\boldsymbol{E}^\text{e}\label{eqn:SVK}\\
\boldsymbol{E}^\text{e} &= \frac{1}{2}\left({\boldsymbol{F}^\text{e}}^\text{T}\boldsymbol{F}^\text{e} - \boldsymbol{1}\right)\\
\boldsymbol{F}^\text{e}(\eta,\boldsymbol{F}) &= \boldsymbol{F}{\boldsymbol{F}^\lambda}^{-1}(\eta)\label{eqn:FeFlam}\\
\boldsymbol{F}^\lambda(\eta) &= \boldsymbol{1}(1-h(\eta)) + \boldsymbol{F}_{\beta^\prime}h(\eta)
\end{align}
\end{subequations}
where $\boldsymbol{1}$ is the second-order isotropic tensor. The components of the matrix phase elasticity tensor $\mathbb{C}^\alpha$ were calculated by Ji and co-workers \cite{jietal2014}, and correspond well with experimental data. The components of the precipitate phase elasticity tensor, $\mathbb{C}^{\beta^\prime}$  were obtained by density functional theory (see Table \ref{tab:elasticity}).

\begin{table}[tb]
    \centering
    \caption{Elasticity constants used for the Mg matrix \cite{jietal2014} and the $\beta^\prime$ precipitate (calculated by Anirudh Natarajan, unpublished data) (GPa).}
    \begin{tabular}{c | c c }
     & Mg ($\alpha$) & $\beta^\prime$\\
    \hline
    $\mathbb{C}_{1111}$ & 62.6 & 78.8\\
    $\mathbb{C}_{2222}$ & 62.6 & 62.9\\
    $\mathbb{C}_{3333}$ & 64.9 & 65.6\\
    $\mathbb{C}_{1122}$ & 26.0 & 24.6\\
    $\mathbb{C}_{2233}$ & 20.9 & 19.9\\
    $\mathbb{C}_{3311}$ & 20.9 & 23.1\\
    $\mathbb{C}_{1212}$ & 18.3 & 11.9\\
    $\mathbb{C}_{2323}$ & 13.3 & 11.6\\
    $\mathbb{C}_{3131}$ & 13.3 & 8.46
    \end{tabular}
    \label{tab:elasticity}
\end{table}

The phase field dynamics are modeled by the diffusion and Allen-Cahn equations, of the following forms \cite{AllenCahn1979,Kim1999}, posed on the reference configuration, $\Omega$:
\begin{subequations}
\begin{alignat}{3}
\frac{\partial c}{\partial t} &= -\nabla\cdot\boldsymbol{J} &&\text{in} &&\Omega\times [0,T]
\label{eq:fick}\\
\frac{\partial \eta}{\partial t} &= -L\mu_\eta &&\text{in} &&\Omega\times [0,T]
\label{eq:AC}\\
\boldsymbol{J}\cdot\boldsymbol{n} &= 0 &&\text{on} &&\partial\Omega\times [0,T]\\
\nabla\eta\boldsymbol{\kappa}\cdot\nabla\eta &= 0 &&\text{on} &&\partial\Omega\times [0,T]\label{eq:ACbc}
\end{alignat}
\end{subequations}
where the flux is defined by $\boldsymbol{J} := -M\nabla\mu_c$, $M$ is the mobility, and $L$ is a kinetic coefficient. The chemical potentials $\mu_c = \delta \Pi/\delta c$ and $\mu_\eta = \delta \Pi/\delta \eta$ are found using standard variational methods, giving the following expressions when assuming $\nabla\eta\cdot\boldsymbol{\kappa}\boldsymbol{n} = 0$ on $\partial \Omega$ (resulting from requiring equilibrium with respect to $\eta$ at the boundary, $\partial\Omega$):
\begin{align}
\mu_c &= \frac{\partial \psi_\text{c}^\alpha}{\partial c}\left(1-h(\eta)\right)+\frac{\partial \psi_\text{c}^{\beta^\prime}}{\partial c}h(\eta)\\
\mu_\eta &= \left[\psi_\text{c}^{\beta^\prime} - \psi_\text{c}^\alpha - \mu_c(c^{\beta^\prime} - c^\alpha)\right]\frac{\partial h}{\partial \eta} - \nabla\cdot\boldsymbol{\kappa}\nabla\eta + \omega\frac{\partial \widetilde{\psi}}{\partial \eta} + \frac{\partial \psi_\text{e}}{\partial \eta}
\end{align}
where
\begin{align}
\frac{\partial \psi}{\partial \eta} &= \left(\frac{1}{2}\boldsymbol{E}:(\mathbb{C}^{\beta^\prime} - \mathbb{C}^\alpha):\boldsymbol{E} - 
\boldsymbol{P}:\left(\boldsymbol{F}^\text{e}(\boldsymbol{F}^{\beta^\prime} -\boldsymbol{1})\boldsymbol{F}^{\lambda^{-1}}\right)\right)\frac{\partial h}{\partial \eta}
\end{align}
The phase field equations (\ref{eq:fick}-\ref{eq:ACbc}) are coupled with nonlinear elasticity posed in terms of the first Piola-Kirchhoff stress, $\boldsymbol{P} = \partial\psi_\text{e}/\partial \boldsymbol{F}^\text{e}$ on the reference configuration $\Omega$: 
\begin{subequations}
\begin{align}
\mathrm{Div}\left(\boldsymbol{P}{\boldsymbol{F}^\lambda}^{-\mathsf{T}}\right) &= \boldsymbol{0} \text{ in } \Omega
\label{eq:elaststrongformgov}\\
\left(\boldsymbol{P}{\boldsymbol{F}^\lambda}^{-\mathsf{T}}\right)\boldsymbol{N} &= \boldsymbol{0} \text{ on } \partial\Omega_{0_T}: \;\text{Neumann boundary conditions}\\
\boldsymbol{u}\cdot\boldsymbol{N}&=0 \text{ on }\partial\Omega_{0_u}: \; \text{Dirichlet boundary conditions}
\label{eq:elaststrongformbc}
\end{align}
\end{subequations}

\subsubsection{Graph generation by states of the binary alloy system without indexing by time}
\begin{figure}[tb]
        \centering
\begin{minipage}[t]{0.24\textwidth}
        \centering
	\includegraphics[width=0.7\textwidth]{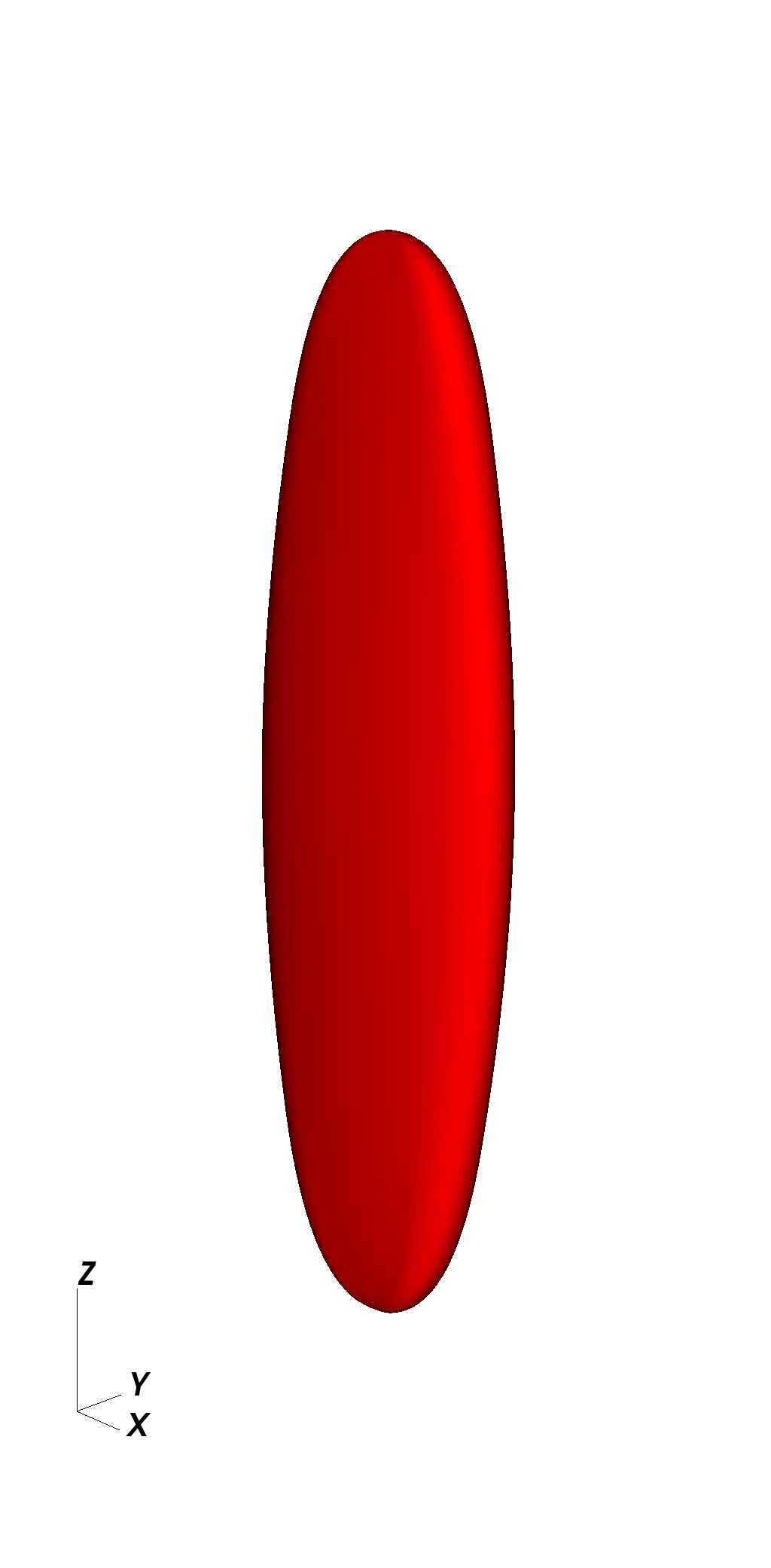}
	\captionof{subfigure}{Iteration 1}
\end{minipage}
\begin{minipage}[t]{0.24\textwidth}
        \centering
	\includegraphics[width=0.7\textwidth]{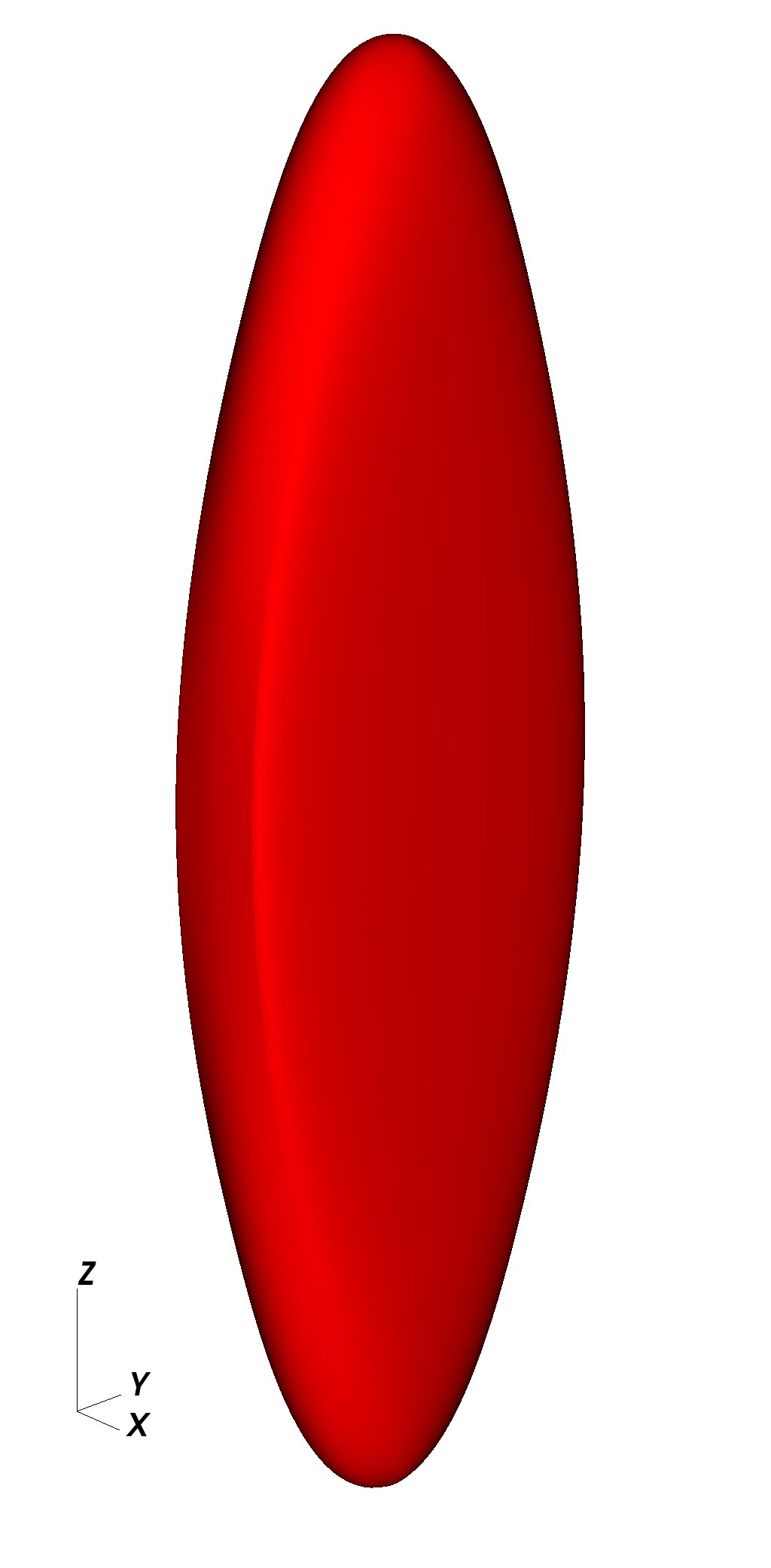}
	\captionof{subfigure}{Iteration 4}
\end{minipage}
\begin{minipage}[t]{0.24\textwidth}
        \centering
	\includegraphics[width=0.7\textwidth]{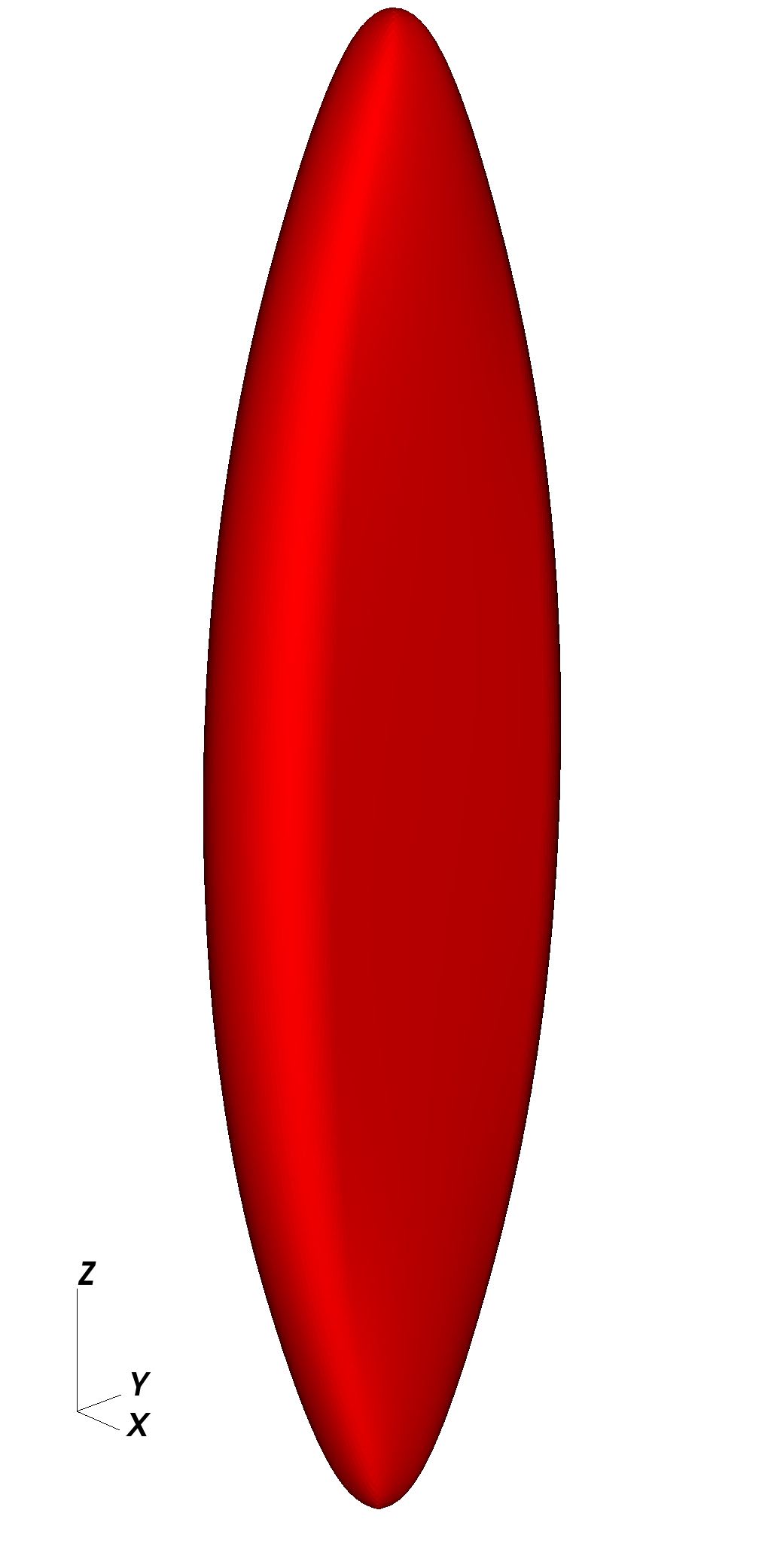}
	\captionof{subfigure}{Iteration 8}
\end{minipage}
\begin{minipage}[t]{0.24\textwidth}
        \centering
	\includegraphics[width=0.7\textwidth]{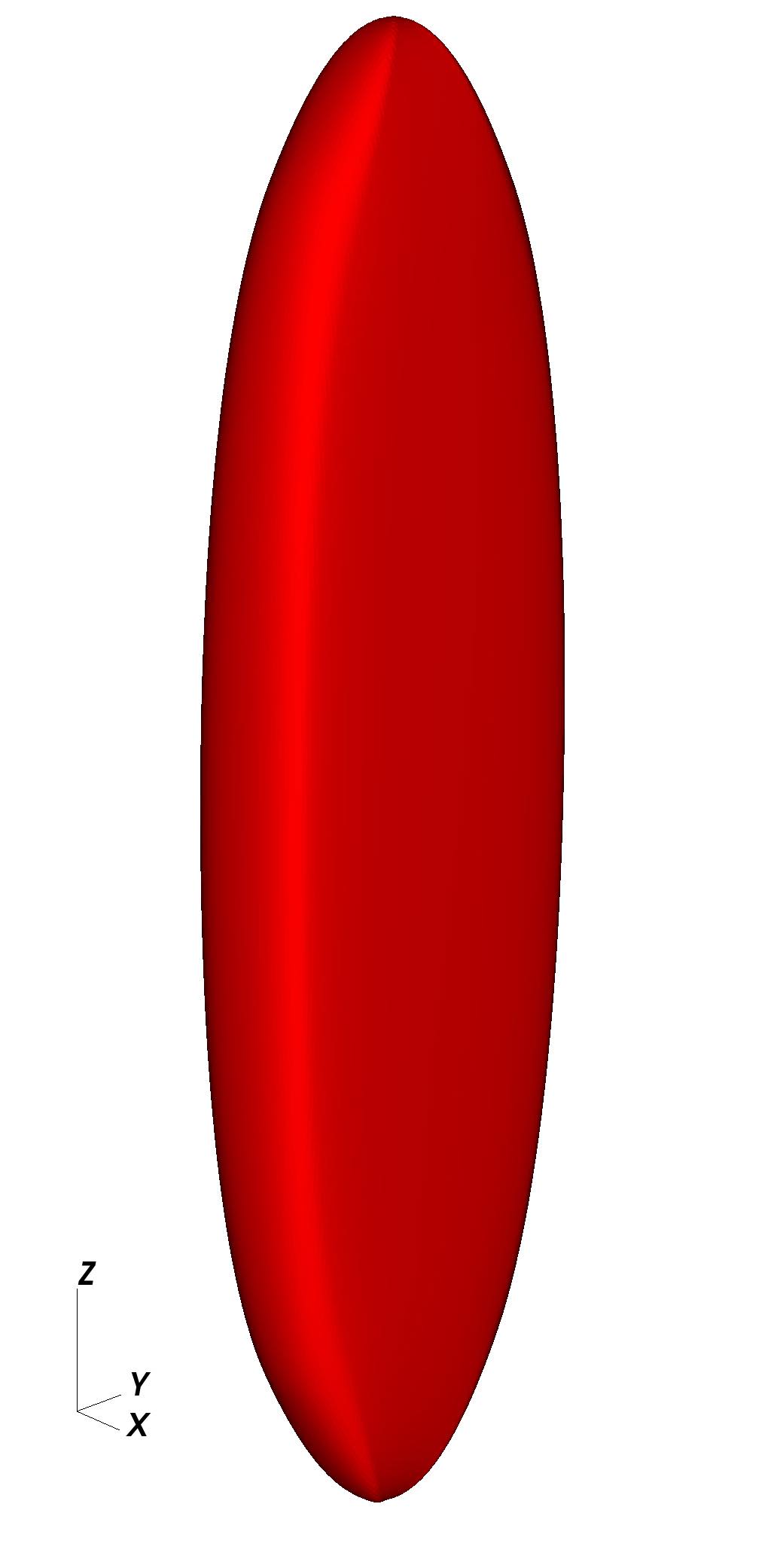}
	\captionof{subfigure}{Iteration 12}
\end{minipage}
        \caption{The  energy-minimizing precipitate geometry at each stage from a sequence that converges toward a local minimum using an approach that draws on machine learning, sensitivity analysis and surrogate optimization.}
	\label{fig:DNS-MLppt}
\end{figure}

Traditionally, studies of precipitate formation, some examples of which have been referred to above \cite{Kim1999,Liu2013,jietal2014,Teichert2018}, have solved the phase field equations (\ref{eq:fick}-\ref{eq:ACbc}) with or without elastic effects (\ref{eq:elaststrongformgov}--\ref{eq:elaststrongformbc}).
Motivated by the search for equilibrium precipitate shapes resulting from phase transformations, we have recently exploited the principle of energy minimization as an alternative to phase field dynamics in a binary alloy system. We define the states of the system as $\mathscr{S}_i = (a_i,b_i,c_i,t_{1_i},\dots,t_{8_i},c_{\text{p}_i},\Pi_i)$. Here, $a_i,b_i$ and $c_i$ define the dimensions of a  rectangular prism that bounds the precipitate, $t_{1_i},\dots,t_{8_i}$ are parameters that define the control points in a spline basis for the precipitate shape, $c_{\text{p}_i}$ is the alloy concentration and as in Sections \ref{sec:nonconvexelasticity} and \ref{sec:CHdynamicsgraphs} $\Pi_i = \int_{\Omega}\psi_i\mathrm{d}V$ is the total free energy of the state. Using direct numerical simulation, we obtain $\sim \mathcal{O}(10^5)$ states of the system for a single precipitate. 
Our approach is to combine machine learning representations, sensitivity analysis and surrogate optimization to find a local minimum of $\Pi$, which corresponds to an equilibrium configuration for the precipitate-matrix system. Figure \ref{fig:DNS-MLppt} shows the energy-minimizing precipitate geometry at four stages from a sequence that converges toward a local minimum using this approach. Details of the methods, convergence to minima, comparisons with experiment and computational cost tradeoffs have been presented elsewhere \cite{Teichert2018}. 

The governing equations \eqref{eq:fick} and \eqref{eq:AC} induce a  dissipative character to the IBVP. For this reason, graphs developed from states computed by these equations supplemented by the elasticity equations (\ref{eq:elaststrongformgov}--\ref{eq:elaststrongformbc}) would be isomorphic to those in Section \ref{sec:CHgraphs} and Figure \ref{fig:dissip2-comp}. In particular, a tree with no branches would result for each IBVP solved. In contrast, the pre-computed states introduced above,  $\mathscr{S}_i = (a_i,b_i,c_i,t_{1_i},\dots,t_{8_i},c_{\text{p}_i},\Pi_i)$, have not been connected by edges defined by a nonlinear time-stepping solution scheme, as was the case in Section \ref{sec:CHgraphs}. This presents other approaches for edge definition by a transition quantity and exploration of the physics via graphs. It is another instantiation of exploration and analysis by graph principles already demonstrated in Section \ref{sec:nonconvexelasticity}.


\subsubsection{Graph completion and exploration principles induced by first-order dissipative dynamics}

\noindent\textbf{Graph chemical potentials and edge definition via a transition quantity}: As a first step, we linearly transform the components $(a_i,\dots,c_{\text{p}_i})$ to each lie in $[0,1]$ and denote this sub-vector as $\boldsymbol{\Xi}_i$. The state can then be written as $\mathscr{S}_i = (\boldsymbol{\Xi}_i,\Pi_i)$. We observe that $\mathscr{S} \in \mathbb{R}^{13}$ is an effective, lower dimensional representation of the state than the spatio-temporal field $\boldsymbol{\zeta} := (c,\eta,\boldsymbol{u})$ in a direct numerical simulation that has $\sim \mathcal{O}(10^6)$ degrees of freedom. Denoting the changes in states between $\mathscr{S}_j$ and $\mathscr{S}_i$ by $\Delta \boldsymbol{\Xi}_{ij} = \boldsymbol{\Xi}_j - \boldsymbol{\Xi}_i$, and $\Delta\Pi_{ij} = \Pi_j - \Pi_i$, the term 
\begin{equation}
\mu_{\Xi_{ij}} := \Delta \Pi_{ij}/\Vert\Delta \boldsymbol{\Xi}_{ij}\Vert
\label{eq:graphchempot1}
\end{equation} 
becomes a generalized chemical potential defined on the graph.

Returning to the task of defining edges between the vertices (states), we note that the principle of maximum dissipation of free energy between states (alternately, a steepest gradient principle) provides one criterion: Given a state $(\boldsymbol{\Xi}_j,\Pi_j)$, we introduce a directed edge to a subsequent state $(\boldsymbol{\Xi}_i,\Pi_i)$ provided  the graph chemical potential, $\mu_{\Xi_{ij}} \le 0$ and $\mathscr{S}_i$ minimizes $\mu_{\Xi_{ij}}$ over pairs $\{\mathscr{S}_j,\mathscr{S}_k\}$. Formally stated,
\begin{equation}
    \text{Given}\;\mathscr{S}_j,\,\exists \mathscr{T}_{ij}\,\text{iff}\, \mu_{\Xi_{ij}} \le 0, \;\text{and}\; \mu_{\Xi_{ij}} = \min\limits_{k} \mu_{\Xi_{kj}}
    \label{eq:graphchempot2}
\end{equation}

\begin{figure}
    \centering
    \includegraphics[width=.8\textwidth]{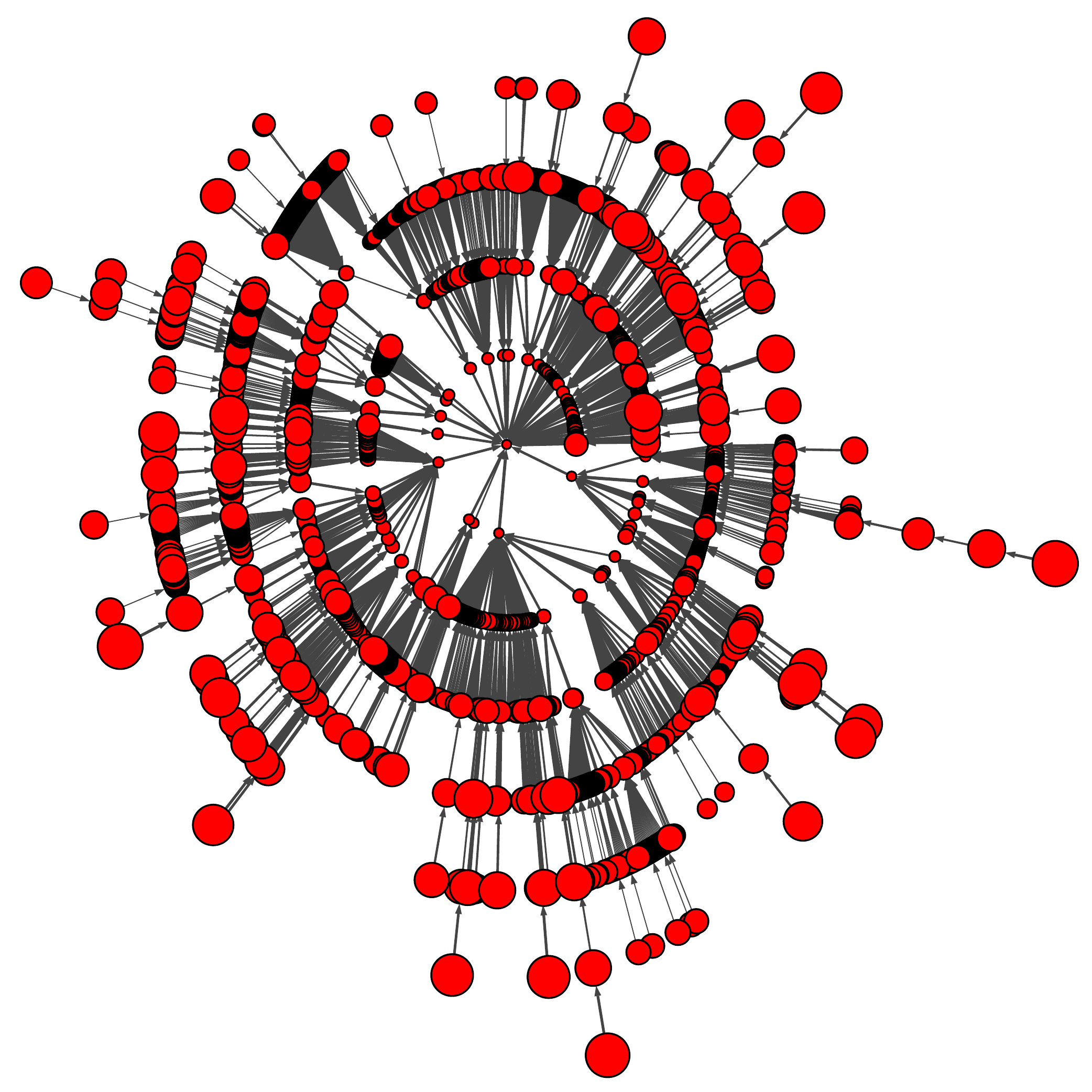}
    \caption{A directed tree graph connecting precipitate states $\mathscr{S}_i = \{\boldsymbol{\Xi}_i,\Pi_i\}$, with edge thickness representing weights defined by Euclidean distance $\Vert\Delta \boldsymbol{\Xi}_{ij}\Vert$, and vertex area proportional to $\log\Pi_i$.}
    \label{fig:ppt_dist1}
\end{figure}

Accordingly, we place a directed edge, $\mathscr{T}_{ij}$ from $\mathscr{S}_j$ to $\mathscr{S}_i$. The graph chemical potential in Equation \eqref{eq:graphchempot1} thus is a transition quantity as discussed in Section \ref{sec:diss-dynamics-properties} and defines edges via Equation \eqref{eq:graphchempot2} with maximum dissipation as an organizing principle. The resulting tree graph is shown in Figure \ref{fig:ppt_dist1} in a circular layout. In this graph, edge weights are defined by Equation \eqref{eq:weight-s} as the Euclidean distance between states $\Vert\Delta \boldsymbol{\Xi}_{ij}\Vert$ and are represented by edge thickness. The vertex areas are proportional to the logarithm of the free energy, $\log\Pi_i$.

\begin{figure}[tb]
        \centering
\begin{minipage}[t]{0.75\textwidth}
        \centering
	\includegraphics[width=\textwidth]{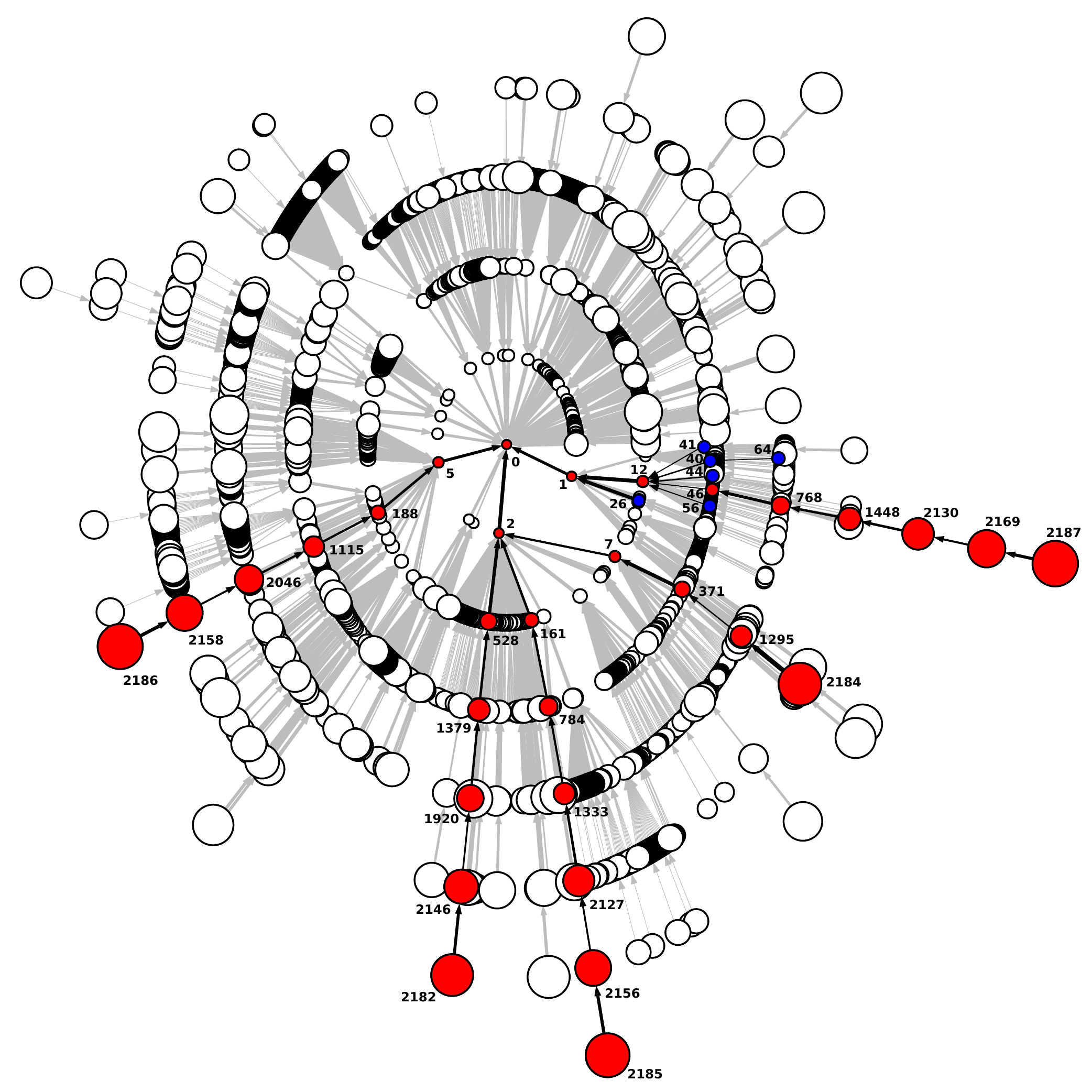}
\end{minipage}%
\begin{minipage}[t]{0.25\textwidth}
        \centering
	\includegraphics[width=\textwidth]{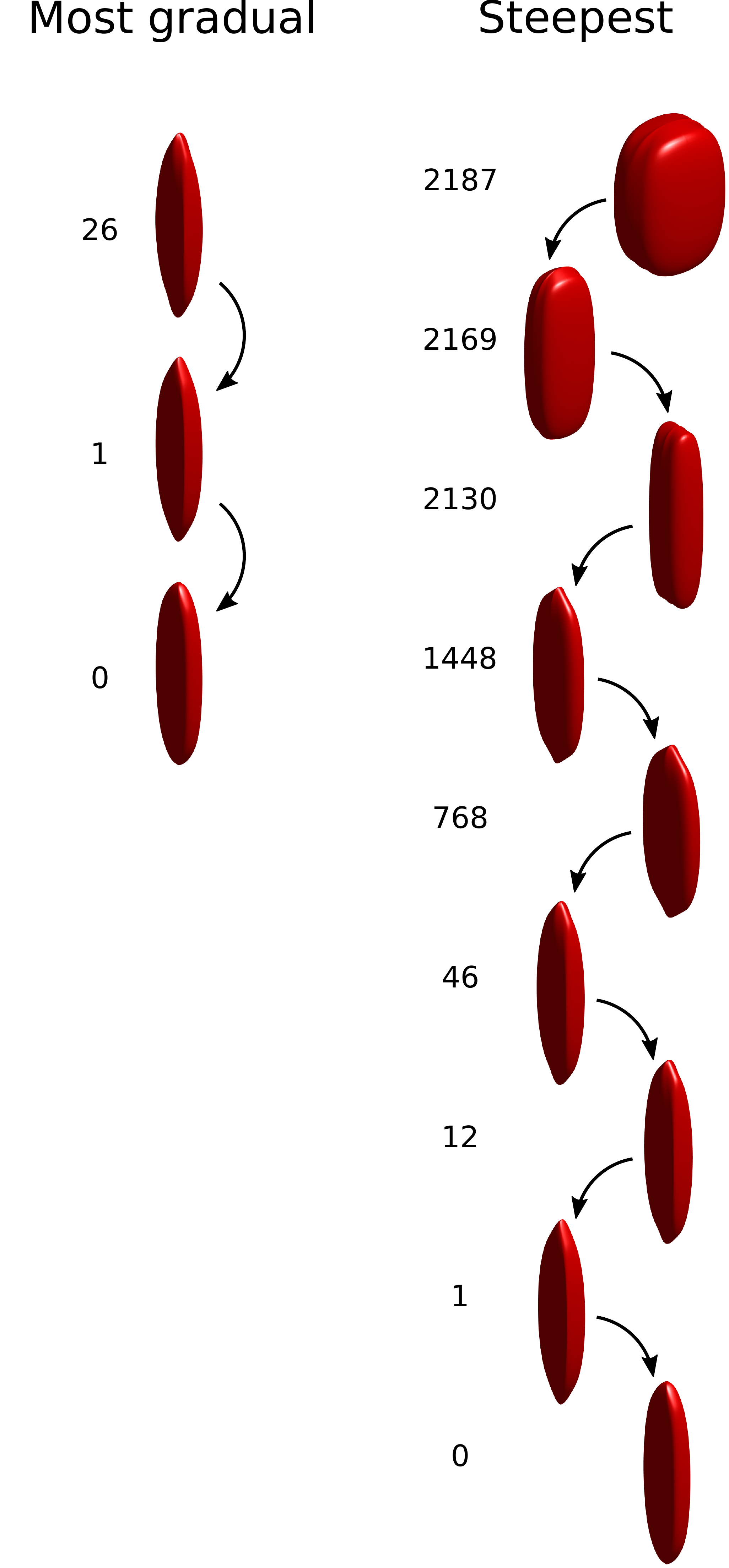}
\end{minipage}
    \caption{The directed tree graph highlighting the five steepest (red) and  five most gradual (blue) energy paths, defined by the magnitude of the (negative) total chemical potential, $\mu_{\Xi_{0k}}$. On the right, note the changes in geometry of the precipitate between states lying along the steepest path, in comparison to the barely discernible changes along the most gradual path. Edge thickness represents weights defined by Euclidean distance $\Vert\Delta \boldsymbol{\Xi}_{ij}\Vert$, and vertex area is proportional to $\log\Pi_i$.}
    \label{fig:ppt_mu}
\end{figure}

\noindent\textbf{Most and least favored energy minimization paths on the graph}: With the graph chemical potential as a transition quantity \eqref{eq:graphchempot1} and maximum dissipation \eqref{eq:graphchempot2} as an organizing principle for definition of edges, the minimum energy state can be identified. In Figure \ref{fig:ppt_mu} it is the vertex corresponding to $\mathscr{S}_0$, the state representing the equilibrium precipitate shape. The remaining vertices are numbered in ascending order of energies, $\Pi_i$. Starting at any vertex corresponding to $\mathscr{S}_i \neq \mathscr{S}_0$, a path is immediately traceable whose each edge represents the maximally dissipative transition among all admissible ones. All paths end in $\mathscr{S}_0$, emphasizing this state's minimum energy property. Paths to $\mathscr{S}_0$ have different numbers of edges--a property that emerges from criterion \eqref{eq:graphchempot2}. Note that this layout was presaged by the graph union $G = G_1\cup G_2$ in Figure \ref{fig:dissipgraph}b.

The five steepest paths between states $\mathscr{S}_k$ and $\mathscr{S}_0$, defined by the magnitude of the (negative) total chemical potential, $\mu_{\Xi_{0k}}$, have been highlighted by red colored vertices. These five paths are identified by their leaf nodes $\{\mathscr{S}_{2182},\mathscr{S}_{2184},\mathscr{S}_{2185},\mathscr{S}_{2186},\mathscr{S}_{2187}\}$ arranged in order of decreasing (increasingly negative) $\mu_{\Xi_{0k}}$. The five most gradual paths by this same criterion are in blue, with leaf nodes $\{\mathscr{S}_{26},\mathscr{S}_{41},\mathscr{S}_{44},\mathscr{S}_{56},\mathscr{S}_{64}\}$ arranged in order of decreasing $\mu_{\Xi_{0k}}$. It also is of interest to note the change in precipitate geometry along the steepest and the most gradual paths. Not surprisingly, there are notable changes between the geometric states along the steepest path, and barely discernible changes along the most gradual path. The edge thickness continues to represent the weights defined by the inter-state Euclidean distances, $\Vert\Delta \boldsymbol{\Xi}_{ij}\Vert$, and the vertex area is proportional to $\log\Pi_i$. Exploration of the graph on the basis of $\mu_\Xi$ thus reveals all admissible, as well as most and least favored paths for energy minimization to the equilibrium precipitate shape.


\begin{figure}[tb]
        \centering
\begin{minipage}[t]{0.8\textwidth}
        \centering
	\includegraphics[width=\textwidth]{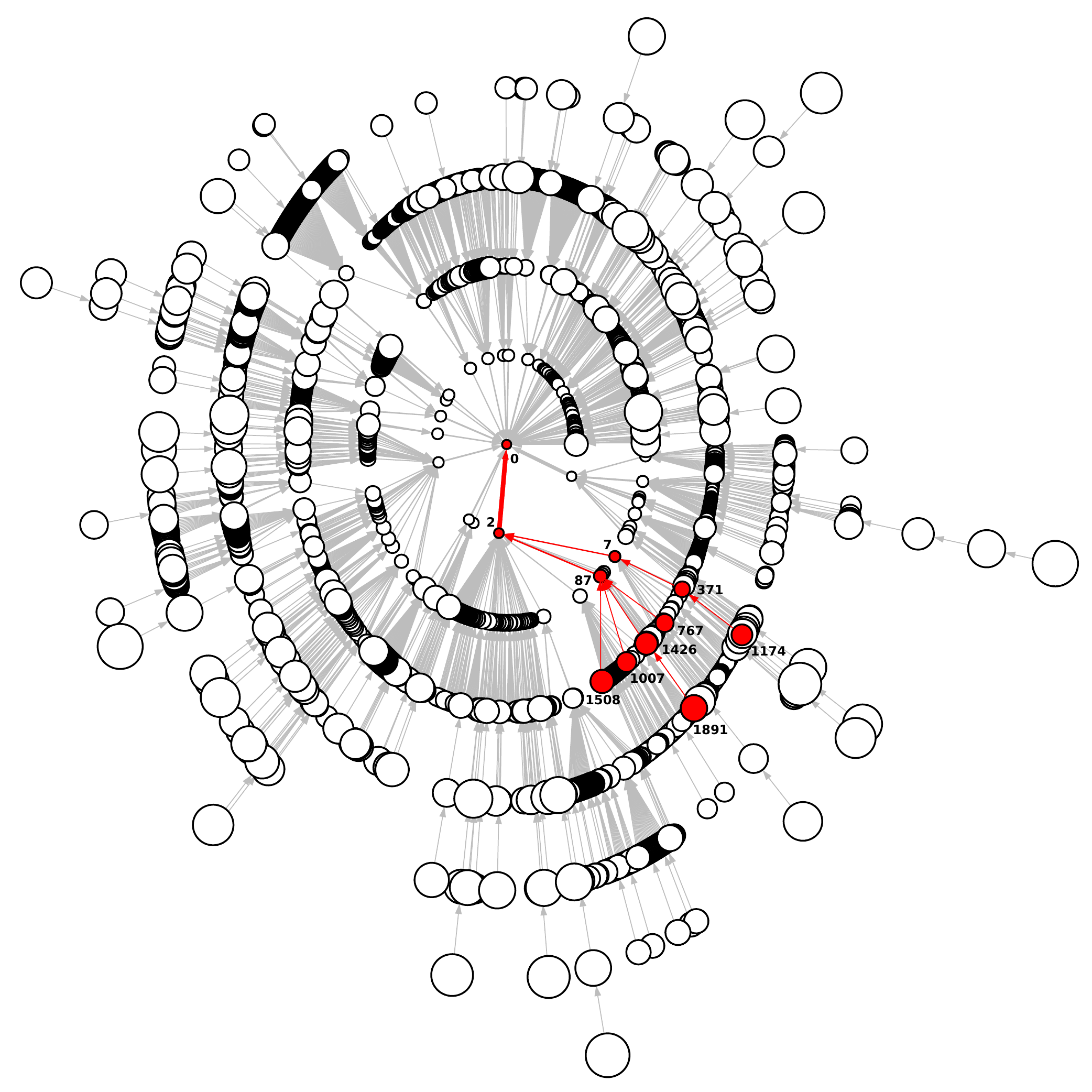}
\end{minipage}%
\begin{minipage}[t]{0.15\textwidth}
        \centering
	\includegraphics[width=\textwidth]{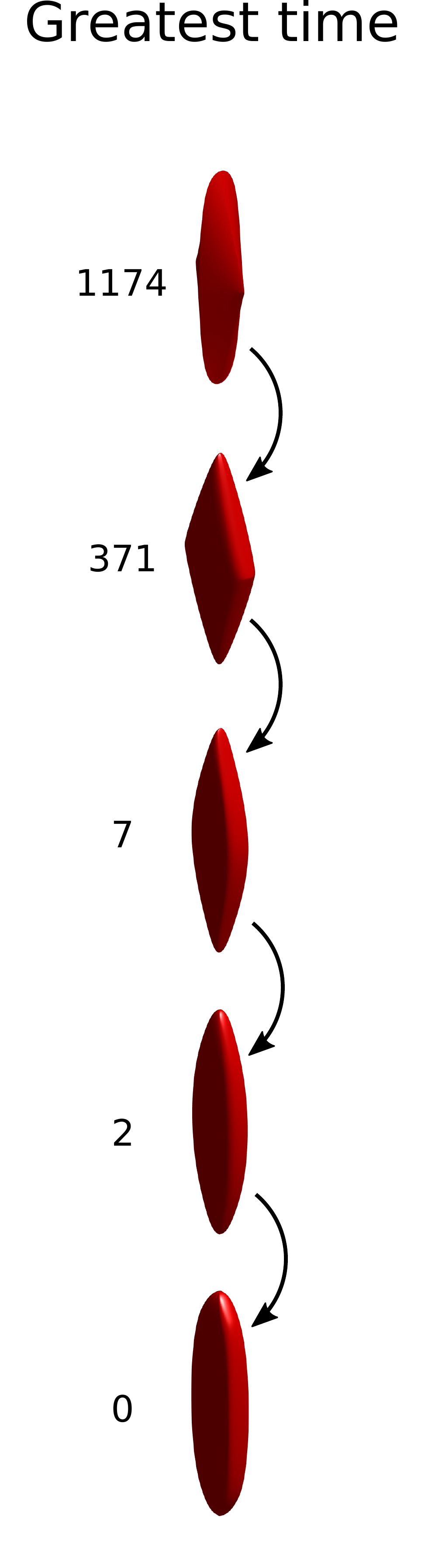}
\end{minipage}
    \caption{The five leaf-to-root paths with the greatest traversal times are plotted with states in red. Here, the edge thickness represents weights defined by graph time of transition of an edge: $\Delta\tau_{ij}$. The path with maximum traversal time is shown with precipitate geometries of the states. Vertex area continues to be scaled by $\log\Pi_i$.}
    \label{fig:ppt_time}
\end{figure}

\noindent\textbf{A graph time for edge transition}: We also introduce a time-like scalar $\tau_i$ for each state $\mathscr{S}_i$, and restrict it to vary such that $\text{sgn}(\Delta\tau_{ij}) = -\text{sgn}(\Delta\Pi_{ij})$. We proceed to arrive at estimates for $\Delta\tau_{ij}$, guided by the graph chemical potential, $\mu_\Xi$. As with $\mu_\Xi$ this is a notion of ``graph time''. Guided by Equation \eqref{eq:AC}, we first observe that we can define an energy gradient-driven analog to the Allen-Cahn equation on the graph with $\Lambda \ge 0$ now denoting a ``kinetic'' coefficient on the graph:

\begin{align}
    \frac{\Vert\Delta \boldsymbol{\Xi}_{ij}\Vert }{\Delta \tau_{ij}} &= -\Lambda\mu_{\Xi_{ij}}\nonumber\\
    \implies \Delta \tau_{ij} &= -\frac{\Vert\Delta \boldsymbol{\Xi}_{ij}\Vert}{\Lambda\mu_{\Xi_{ij}}}\nonumber
\end{align}
Recalling $\mu_{\Xi_{ij}} = \Delta \Pi_{ij}/\Vert\Delta \boldsymbol{\Xi}_{ij}\Vert$, where $\Delta \Pi_{ij} \le 0$ we arrive at 
\begin{equation}
\Delta \tau_{ij} \sim \frac{\Vert\Delta\boldsymbol{\Xi}_{ij}\Vert^2}{\vert\Delta\Pi_{ij}\vert}
\label{eq:statetime}
\end{equation}
requiring, of course, that $\Delta\Pi_{ij} \neq 0$. Equation \eqref{eq:statetime} suggests that the time to traverse an edge is related to the squared ``distance'' between states in $\mathbb{R}^{12}$ scaled by the magnitude of the corresponding energy decrease. This is in agreement with the physics of first-order kinetic processes according to which a rate rises with greater energy decreases, or the transition time decreases. Conversely, an increased ``distance'' between states in $\mathbb{R}^{12}$ decreases the rate, increasing the transition time. With this measure of graph time, we can explore the time required to traverse a leaf-to-root path. Figure \ref{fig:ppt_time} shows the five such paths with greatest time of traversal. It is notable that these are distinct from the five most gradual energy paths in Figure \ref{fig:ppt_mu}: The cumulative effect  of transitions across each edge determines this traversal time, rather than the average steepness of the paths. Also shown are the changes in precipitate geometry along the path with the maximum time of traversal. Following Equation \eqref{eq:statetime}, large transition times result from larger changes in geometry, $\Vert\Delta \boldsymbol{\Xi}_{ij}\Vert$, over the first two edges, followed by small energy changes, $\Delta\Pi_{ij}$, over the last two edges. Also note that the weights have now been defined by graph time of transition via Equation \eqref{eq:weight-t}, and are represented by edge thickness in Figure \ref{fig:ppt_time}.


\begin{figure}
    \centering
    \includegraphics[width=.8\textwidth]{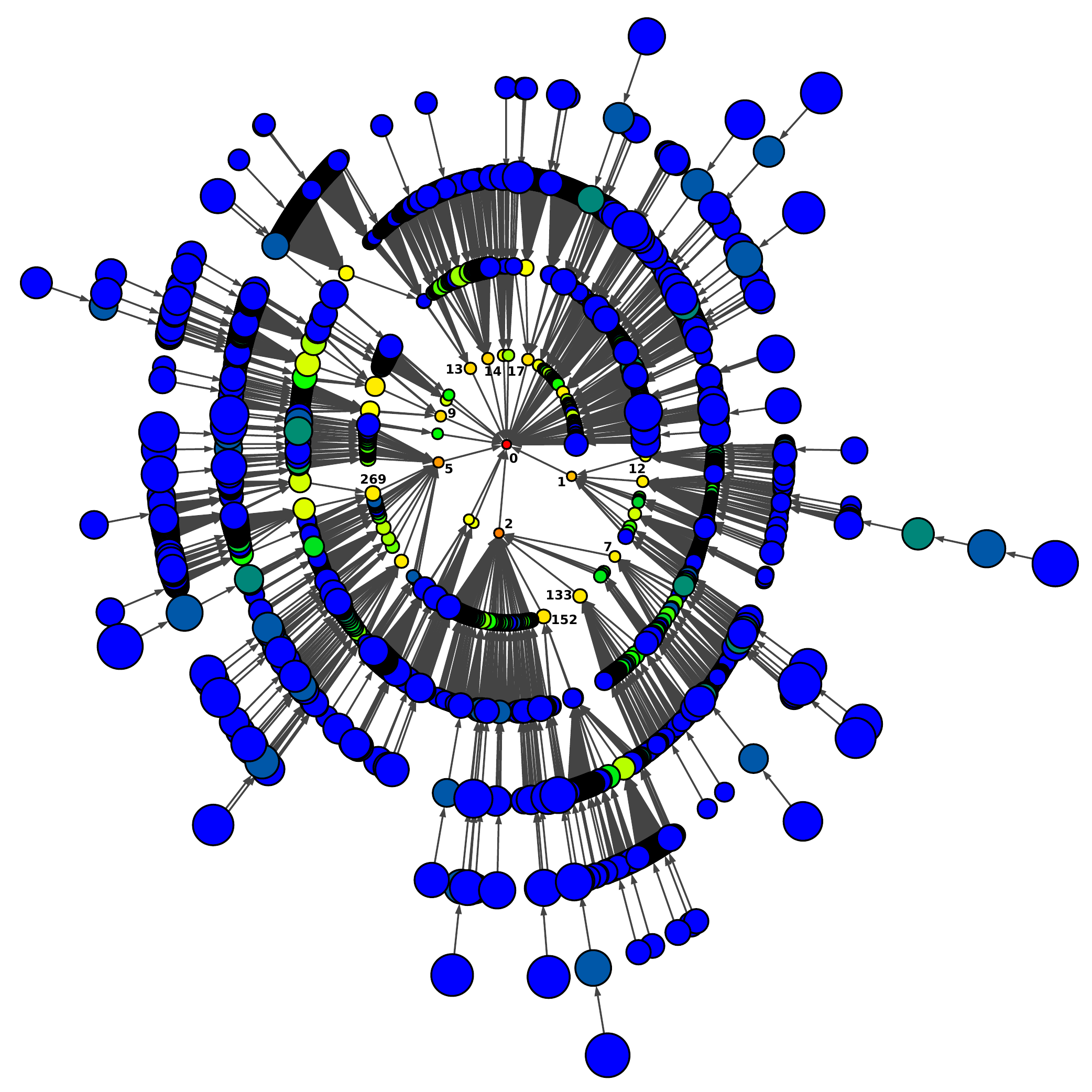}
    \caption{The directed tree graph with vertices shaded according to the log of the PageRank, where red is the highest PageRank and blue is the lowest. Edge thickness represents weights defined by distance $\Vert\Delta \boldsymbol{\Xi}_{ij}\Vert$, and vertex area is proportional to $\log\Pi_i$.}
    \label{fig:ppt_pagerank}
\end{figure}

\noindent\textbf{PageRank as a measure of the relative importance of states for energy minimization paths}: The final aspect that we study is centrality of the vertices representing states. The motivation is quite clear: As the graph illustrates in Figures \ref{fig:ppt_mu} and \ref{fig:ppt_time}, the vertex representing state $\mathscr{S}_0$ has the highest in-degree, while this measure decreases with increasing neighbor separation from it. Figure \ref{fig:ppt_pagerank} uses the PageRank of the corresponding vertices \cite{Newman2010}, to show the relative importance of states in enabling the dynamic process of energy minimization. Since this measure scales the in-degree by the out-degree, it indicates the states that pool many others while allowing paths to fewer downstream states, thus ``focusing'' the flow of the dynamic process of energy minimization. The weights are defined by the Euclidean distance between states and are represented by edge thickness. The areas of the vertices are proportional to $\log\Pi_i$.


\section{Closing remarks}
\label{sec:closingremarks}

We make the case that graph theory offers a framework for representation, exploration and analysis of large scale computed solutions. The fundamental insight required is that high-dimensional field solutions typically admit functional representations of low-dimensional states, which are the vertices of a graph. Transitions between states, which could either be a (nonlinear) solution step, or be defined by the change in a physical property, are the edges of the graph. With this foundation, isomorphisms can be identified between the computational and physical framework of states and transitions on one hand, and graph vertices and edges on the other. In this isomorphism, many properties of the numerical solution procedure and of the dynamical physical system are in correspondence with properties of graphs. This includes standard notions of weights, directedness of edges, connectedness of graphs, components, cliques and cycles. Other correspondences arise when considering specific dynamical systems as graphs. 

The framework-setting observations that have been summarized above are rendered concretely in considering four systems. In this regard, the nature of the graphs constructed on non-dissipative elastodynamics and linear elasticity are simple. The approach becomes more profound for the graphs constructed on states of gradient-regularized, non-convex elasticity, and graphs on first-order dissipative dynamical systems with and without time-series data. Here, the reach of the graph theoretic approach is apparent in its introduction of a framework to organize these solutions, and then to explore and analyze them. Some observations are useful in this setting:
\begin{itemize}
    \item Comparing the three more consequential cases that were recalled above, we note that perhaps the greatest utility of the graph theoretic approach occurs where the data set of solutions needs an organizing principle to be imposed on it. This was the case with the nonlinear solution step inducing edges for the graph of strain states, and the maximum dissipation principle doing the same for the graph on states of the binary alloy system. These graphs are rich in having many branches, and interesting paths, cliques and cycles. Notably, the graph theoretic framework provides insight to the physics of these systems in terms of accessibility of strain states, paths between them, energy minimizing paths, and notions of graph chemical potentials, graph time and several others. In contrast, the graph on the three-phase transforming system indexed by time naturally has a linear, tree graph without branches. In this case, the strict condition of the second law of thermodynamics imposes directedness and barriers to arbitrary introduction of edges (transitions) between vertices (states).
    \item It is particularly insightful to note how two physical systems, both of which would be solved by imposing first-order dynamics in the traditional computational physics approaches, lead to very different graphs if the dynamics are eliminated in one case. Here, the graph constructed on states of the binary alloy system using maximum dissipation to induce edges between states reveals a rich array of possible paths to energy minimization, which yield to exploration and analysis. In contrast the time series information in the graph on the three phase transforming system has a single path and less room for further exploration.
    \item It is important to observe that the notion of low-dimensional states underlying the framework here is quite different from dimensionality reduction as it works in proper orthogonal decomposition, tensor decomposition, multi-fidelity modelling and related methods. High-fidelity, and therefore high-dimensional, \emph{field solutions} are assumed here (although the approach works for low-dimensional field solutions, also). The low-dimensional \emph{functionals} that form states are extracted from the high-fidelity, high-dimensional field solutions. These functionals and states are natural, ``aggregate'' physical quantities relevant to the dynamical physical system. In addition to the average strain, phase volume fractions, various global energy measures, precipitate shape parameters and average compositions employed here, we recall a few others: lift, drag and thrust in fluid dynamics, load at yield and failure strain in solid mechanics, charge state and voltage in electro-chemistry, and the magnetic moment in electro-magnetism.
\end{itemize}

It will be instructive to extend these graph theoretic ideas, and explore examples more deeply than the preliminary consideration of stationary or dynamical, dissipative or non-dissipative systems in this first communication. Such a broader framework would bring the graph theoretic perspective of computational physics closer to decision science for natural and engineered systems.

\section*{Acknowledgements}
We gratefully acknowledge the support of Toyota Research Institute, Award \#849910, ``Computational framework for data-driven, predictive, multi-scale and multi-physics modeling of battery materials"; NSF DMREF grant: DMR1436154, ``DMREF: Integrated Computational Framework for Designing Dynamically Controlled Alloy-Oxide Heterostructures"; Sandia National Laboratories via its LDRD mechanism: 746300, ``Material Variability Research''; and the U.S. Department of Energy, Office of Basic Energy Sciences, Division of Materials Sciences and Engineering under Award \#DE-SC0008637 that funds the PRedictive Integrated Structural Materials Science (PRISMS) Center at University of Michigan.  Computing resources were provided in part by the NSF via grant 1531752 MRI: Acquisition of Conflux, A Novel Platform for Data-Driven Computational Physics (Tech. Monitor: Ed Walker). This work also used the Extreme Science and Engineering Discovery Environment (XSEDE) Comet at the San Diego Supercomputer Center and Stampede2 at the Texas Advance Computing Center through allocations TG-MSS160003 and TG-DMR180072.
\bibliography{refs}
\end{document}